\documentclass[journal=mamobx,manuscript=reprint]{achemso}
\SectionNumbersOn
\usepackage{subfiles}
\usepackage{xr}
\usepackage{chemformula} 
\usepackage[T1]{fontenc} 
\usepackage[utf8]{inputenc} 

\usepackage{float}
\usepackage{amsmath}
\usepackage{subfigure}
\usepackage{cellspace}
\usepackage{csquotes}
\usepackage{multirow}
\usepackage{booktabs}
\usepackage{gensymb}
\usepackage{amssymb}
\usepackage{wasysym}

\author{Atmika Bhardwaj}
\affiliation{Theory of Polymers, Leibniz-Institut für Polymerforschung Dresden e.V., Hohe Strasse 6, 01069 Dresden, Germany.}
\alsoaffiliation{Institute of Theoretical Physics, Technische Universität Dresden, Zellescher Weg 17, 01069 Dresden, Germany.}
\alsoaffiliation{These authors contributed equally to this work.}

\author{Huzaifa Shabbir}
\affiliation{Theory of Polymers, Leibniz-Institut für Polymerforschung Dresden e.V., Hohe Strasse 6, 01069 Dresden, Germany.}
\alsoaffiliation{These authors contributed equally to this work.}

\author{Jens-Uwe Sommer}
\email{sommer@ipfdd.de}
\affiliation{Theory of Polymers, Leibniz-Institut für Polymerforschung Dresden e.V., Hohe Strasse 6, 01069 Dresden, Germany.}
\alsoaffiliation{Institute of Theoretical Physics, Technische Universität Dresden, Zellescher Weg 17, 01069 Dresden, Germany.}

\author{Marco Werner}
\email{werner-marco@ipfdd.de}
\affiliation{Theory of Polymers, Leibniz-Institut für Polymerforschung Dresden e.V., Hohe Strasse 6, 01069 Dresden, Germany.}

\title{The Influence of Crosslinking and Deformation on Polymer Crystallization and Melting: A Molecular Dynamics Study}

\begin{document}
\maketitle

\begin{abstract}
We investigate the crystallization of crosslinked and entangled polymers under external deformation using a coarse-grained poly(vinyl alcohol) (CG-PVA) model and molecular dynamics simulations. Following uniaxial deformation, the systems are cooled at a constant rate to form semi-crystalline states and subsequently heated at a constant rate to induce melting. For unstretched systems, network junctions do not significantly affect the nucleation temperature but increase the amorphous fraction and reduce the melting temperature. Uniaxial deformation accelerates nucleation and markedly increases the crystallization temperature, with more strongly crosslinked polymers exhibiting larger shifts that correlate with an enhanced orientation order parameter. We further compare cooling and heating cycles under constant-strain and constant-stress conditions. Under constant stress, crystallization induces additional elongation beyond the initial pre-stretch and leads to pronounced mechanical hysteresis upon heating, a behavior characteristic of reversible shape-memory materials.
\end{abstract}

\section{Introduction}\label{sec:Intro}

Crystallization and melting characteristics of polymers are crucial for the mechanical and thermal properties of practically all polymer-based bulk materials.
During crystallization, partial alignment of molecular chains occurs, transforming them from a random coil (high entropy state) to partially folded contours (low entropy state), leading to thin lamellar structures on the nanometer scale. 
These crystalline domains coexist with amorphous regions to form semi-crystalline polymers~\cite{keller1957note, armitstead1992polymer, strobl1997physics, sommer:piuopc:2007}.
A significant gap between the crystallization and melting temperatures indicates that semi-crystalline polymers is matter out-of equilibrium  which causes an intricate relationship between thermal and mechanical history and final material properties. It is well recognized that entanglements between long polymers chains control nucleation behavior and semi-crystalline morphologies~\cite{luoEntanglements2016, luoRole2016, zou2022molecular}.

As a consequence of the long lifetime of entanglements, polymer crystallization is significantly affected by thermo-mechanical history and can be tuned by varying processing conditions such as cooling rate, crystallization temperature, and cooling methods (isothermal, continuous, or interrupted) \cite{shen2016polymer}. Additionally, polymer architecture in terms of side groups~\cite{alamo:jpc:1984}, branches~\cite{ungar:cr:2001}, rings~\cite{zardalidis:sm:2016}, crosslinks, and poly-dispersity, essentially influences crystallization behavior and the resulting morphologies.

It is a long-known phenomenon~\cite{smith:jrnbsn:1938,gent:tfs:1954,tobolsky:rct:1956,mandelkern:jacs:1959} that the application of deformation strain during cooling affect the memory and mechanical response of polymer-based semi-crystalline materials. In this case, the hysteresis of stress or deformation as a function of temperature can be used to program shape changes.
For example, studies on crosslinked poly-($\epsilon$-caprolactone) cooled under stretched conditions have revealed a drop in mechanical stress during cooling beyond the transition point at a given uniaxial deformation and complementary display a jump in external deformation if stress is fixed~\cite{murcia:m:2021, posada-murcia:m:2022}. These observations qualitatively agree with Flory's original argument~\cite{flory1962morphology}, which suggests that stress in the amorphous part is released as widening stems extend in the stretching direction~\cite{dolynchuk2017reversible}. The argument was developed to explain equivalent phenomena in form of a decay of stress~\cite{gent:tfs:1954,tobolsky:rct:1956,kim:jpsp-pp:1968} and spontaneous elongation~\cite{smith:jrnbsn:1938,mandelkern:jacs:1959} observed in natural rubber~\cite{smith:jrnbsn:1938,gent:tfs:1954,tobolsky:rct:1956,kim:jpsp-pp:1968} and polyethylene~\cite{mandelkern:jacs:1959} upon phase transition.

While there is already a significant number of simulation studies investigating the crystallization behavior in polymer melts~\cite{meyer2001formation, hu2001chain, yamamoto2004molecular, CH_coex2009, luoRole2016, nicholson2020flow, PAYAL2021, bhardwaj2024}, the crystallization of crosslinked polymers has received much less attention.
Experimentally at low temperatures, the impact of crosslinks on crystallization in undeformed materials is well known as a retention for crystallite growth~\cite{LAMBERT1994}. We note that the crystallization of crosslinked polymers is particularly interesting from the perspective of applying external deformation: In contrast to flow-induced crystallization~\cite{WANG2016}, the deformation is applied under quasi-static conditions, which allows us to investigate the impact of external strain directly. At high temperatures, crosslinks lead to a finite elastic modulus while still displaying viscous liquid-like properties which makes elastomers an outstanding material class with many applications in technology and everyday life.
However, their interplay with nucleation and growth in case of deformed materials has not yet been elucidated via computer simulations.

In this work, we investigate the consequences of combining strain and crosslinking on the crystallization in a supercooled system using large-scale molecular dynamics (MD) simulations.
A multi-step protocol, described in Section \ref{sec:methods}, is employed to investigate the influence of crosslinking and deformation in coarse-grained (CG) poly(vinyl alcohol) (PVA).
The protocol includes the following steps: creating the polymer configuration in the melt, initializing crosslinks in the melt state, deforming the samples in the amorphous state, relaxing the samples in various ensembles, and performing cooling and reheating cycles.
Properties such as specific volume, crystallinity, stem length distribution, orientation order, and stress are studied as functions of temperature in Section \ref{sec:results} to provide insights into both microscopic and macroscopic changes occurring during polymer crystallization. 
Furthermore, we study the crystallization and melting of strained samples under constant mechanical load obtaining the characteristic temperature-elongation hysteresis as observed in experiments. We reveal a qualitative difference in crystalline morphology comparing the scenarios under constant elongation and constant load respectively by investigating the orientation order in both cases. 
We summarize the discussions in our conclusions in Section \ref{sec:conclusion}.

\section{Methods}\label{sec:methods}

We employ a CG bead-spring model to simulate PVA, where each CG bead represents a PVA monomer.
The model was simulated using the LAMMPS package~\cite{LAMMPS}, and the force field was aligned with the previously defined model~\cite{meyer2002formation, meyer2001formation, vettorel2007, CH_lammps2009}.
The covalent interactions between neighboring beads in the CG-PVA model are described by bond stretching and bond bending potentials.
A harmonic potential with a spring constant of $\mathrm{k}/2=1.0\times 10^3~\mathrm{k_BT_0/\sigma^2}$ is used for stretching contributions, while a non-trivial tabulated potential is employed for bending contributions.
The use of a tabulated potential is crucial for accurately mapping the torsional potential of the backbone through the bending potential in the CG-PVA model.

\begin{figure}[!ht]
  \begin{center}
  \includegraphics[width=\hsize]{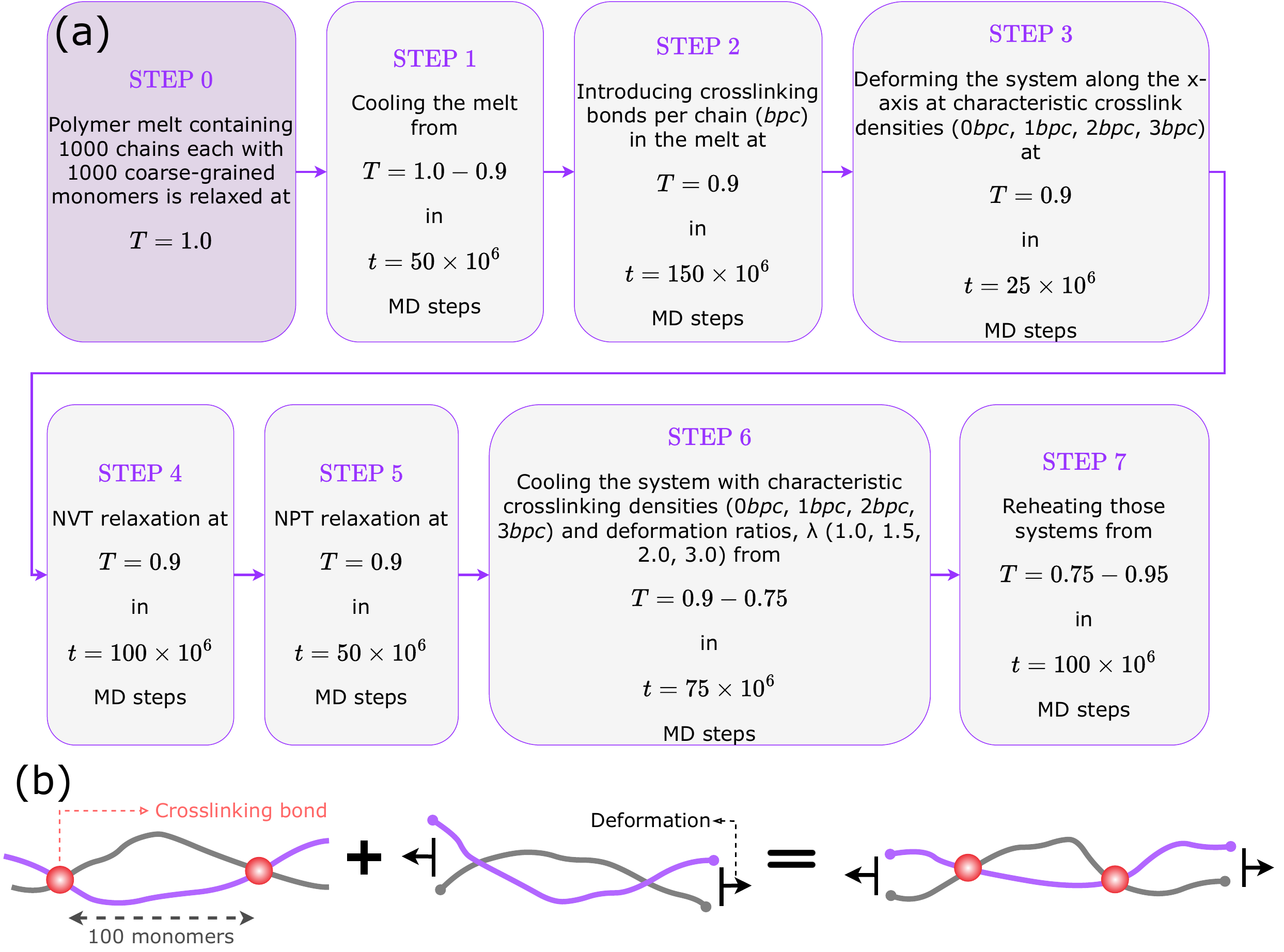}
  \end{center}
  \caption{(a) A summary flowchart of the steps involved in the protocol.
  Step $0$ includes the melt preparation and relaxation.
  The simulation results of each of these steps can be found in the following figures.
  Please refer Figure \ref{fig:protocol_cl} for step $2$, Figure \ref{fig:protocol_deform} for step $3$, SI Figure \ref{fig:nvt} for step $4$, SI Figure~\ref{fig:npt} for step $5$ and SI Figures \ref{fig:pressure} and \ref{fig:box_len} for steps $6$ and $7$, respectively.
  (b) An illustration (not to scale) of two crosslinks (left) and the deformation of two entangled polymer chains along the x-axis (middle).
  A combination of the crosslinking and deformation processes is drawn on the right.}
  \label{fig:flowchart_protocol}
\end{figure}

The protocol employed to investigate the impact of crosslinks and deformation on polymer crystallization and melting is shown in Figure~\ref{fig:flowchart_protocol}.
We adopt the polymer melt configuration from our previous work~\cite{bhardwaj2024}, which consists of $n_{chain} = 1000$ chains, with each chain containing $N = 1000$ CG-PVA monomers or beads, resulting in a total of $10^{6}$ beads.
The initial melt state was prepared by self-avoiding random walks and was then relaxed in NVT at a temperature of $T=1.0$~\cite{CH_md2010}.
The reduced temperature of \(\mathrm{T_0} = 1\) corresponds to \(550 ~ \mathrm{K}\)~\cite{CH_lammps2009}.
After relaxation, the polymer is cooled from $T = 1.0 ~(~\widehat{=}~550 ~\mathrm{K})$ to $T=0.90~(~\widehat{=}~495 ~\mathrm{K})$ in NPT using $50 \times 10^6 $ MD steps.
This is done to lower the computational costs of running each cooling cycle (discussed later in this section) by almost $40\%$.
This temperature, $T=0.9$ is well above the melting point of an unstrained sample.
The cooling rate is $1.9\times 10^{-7}\tau^{-1} ~(~\widehat{=}~43 ~\mathrm{K}/ \mu\mathrm{s})$, with a time step of MD integration as $0.01\tau ~(~\widehat{=}~25.5 ~\mathrm{fs})$.

\begin{figure}[!ht]
  \subfigure[]
    {
    \includegraphics[width=0.45\hsize]{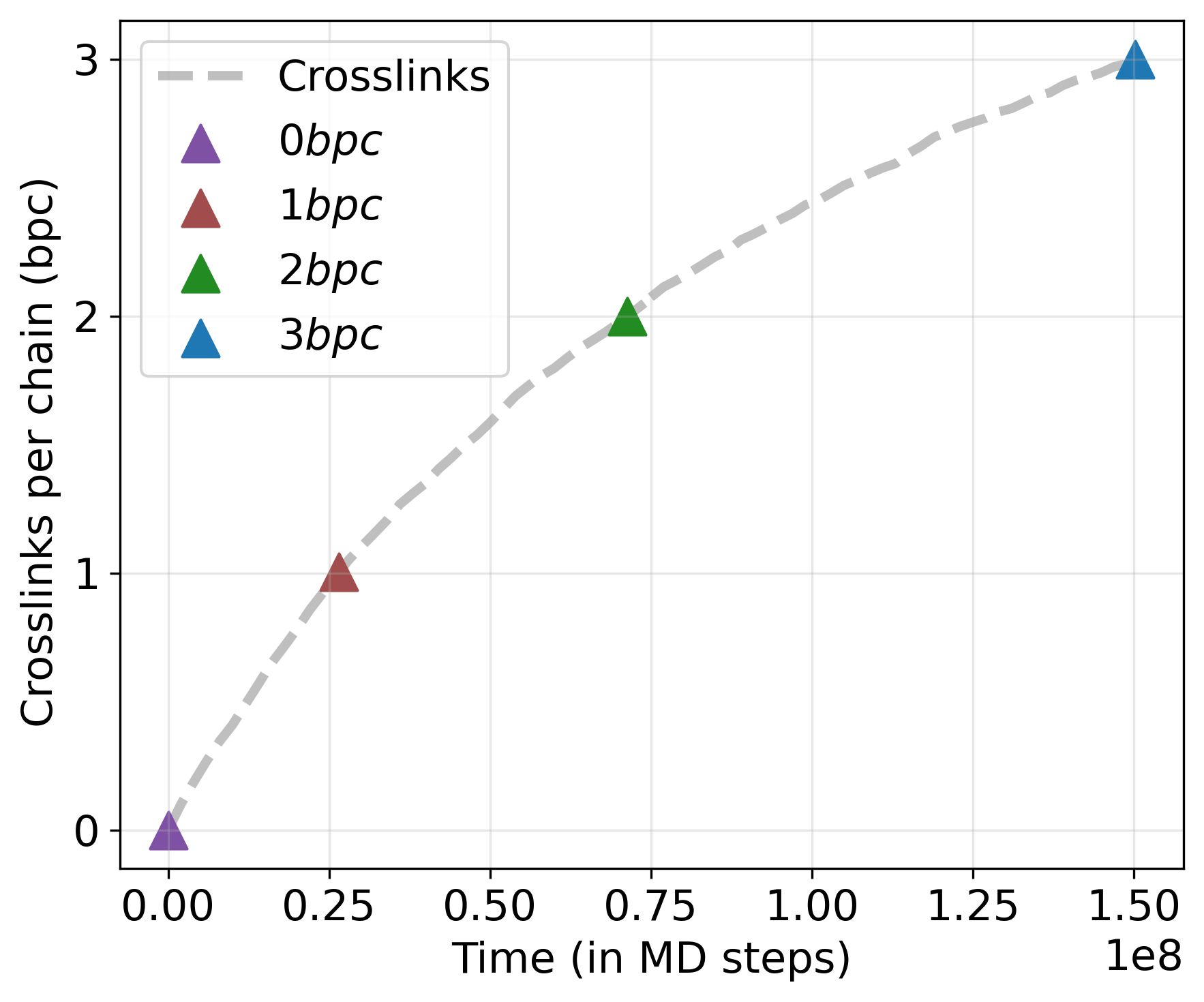}
    \label{fig:protocol_cl}
    }
  \subfigure[]
    {
    \includegraphics[width=0.45\hsize]{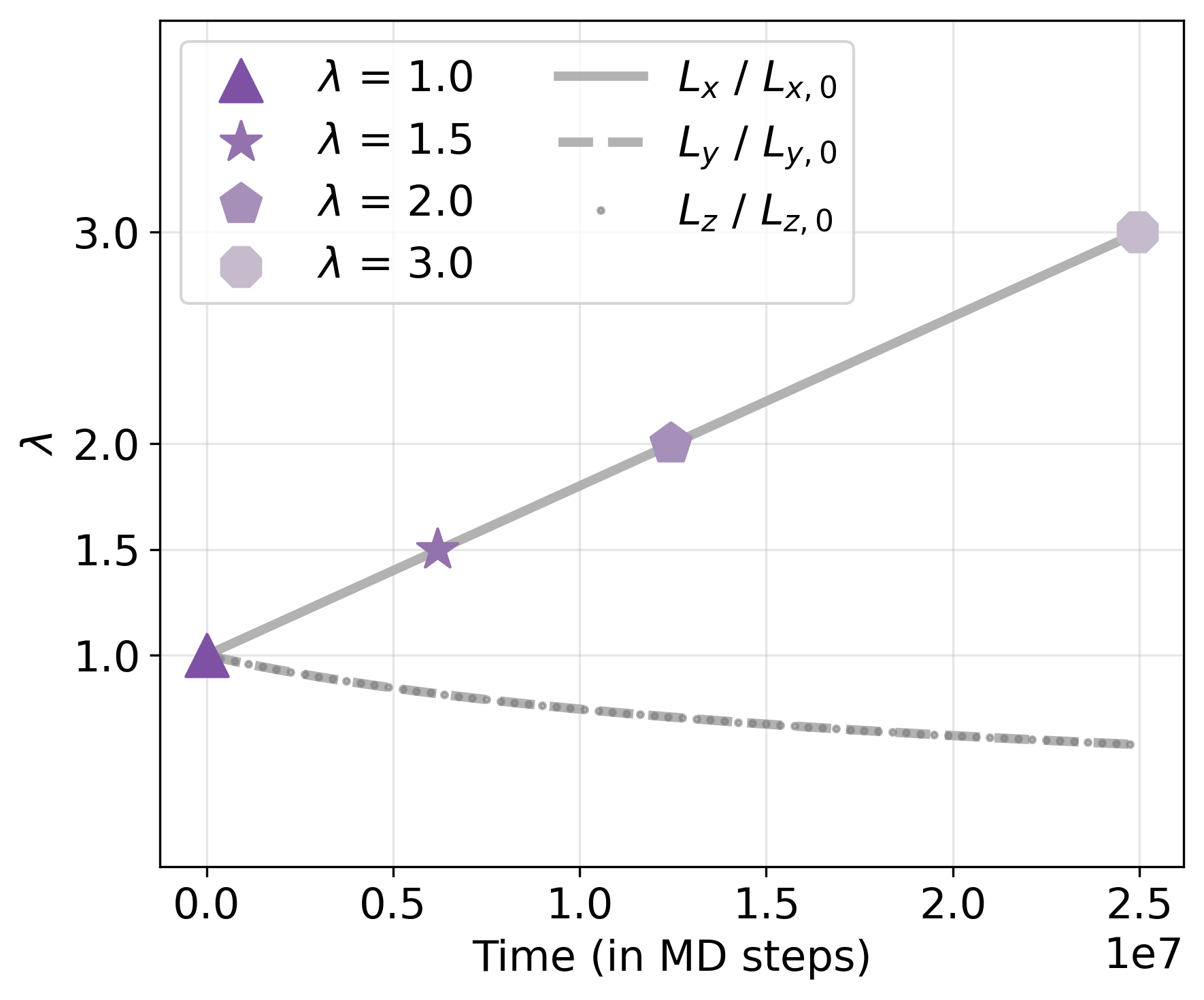}
    \label{fig:protocol_deform}
    }
  \caption{Illustrations of the crosslinking and deformation steps involved in the protocol.
  (a) Starting with a relaxed melt state, we introduce crosslinks in the system.
  The dashed line marks the evolution of crosslinks, whereas the colored upright triangles denote the characteristic configurations of average bonds per chain chosen to apply the deformation in the next step.
  (b) Following the process of crosslinking, the system is uniaxially deformed along the x-axis,~$\lambda$.
  The system with $\lambda=1.0$ is the undeformed system, refer to Equation \ref{eq:lambda}.
  Deformation steps continue until $\lambda$ reaches $3.0$.
  }
  \label{fig:protocol_cl_deform}
\end{figure}

In the second step of the protocol, we introduce crosslinks in the melt state at $T=0.90$.
This is performed in the NPT ensemble to embed local effective monomer volume changes due to the crosslinking.
We tag every $100$th monomer on each chain as a crosslinkable site.
Permanent crosslinks are then established upon pair contact between two crosslinkable monomers.
We plot the grey-dashed line to visualize the evolution of the average number of crosslinking bonds per chain (bpc) in Figure \ref{fig:protocol_cl}.
The points highlighted at $0bpc,~1bpc,~2bpc, ~\text{and}~3bpc$ represent the configurations selected to apply the deformation in the next step.

The third protocol step performs uniaxial deformation due to stretching along the x-axis in the NVT ensemble at constant $T= 0.90$.
The process, as shown in Figure \ref{fig:protocol_deform}, is simulated in $25\times 10^6$ MD steps.
We apply uniaxial deformation, where the deformation ratio is defined with respect to the $x-$axis as denoted by $\lambda$ such that
\begin{equation}
\lambda = L_{x}/L_{x,0}~~.
\label{eq:lambda}
\end{equation}
Here, $L_x$ ($L_{x,0}$) is the dimension of the deformed (undeformed) box.
Since volume is conserved in dry polymers under external deformation, the dimensions along the remaining axis follow as $L_{y/z}=L_{y/z,0}/\sqrt{\lambda}$.
During the deformation process, we select deformation ratios $\lambda$ that are highlighted in the legend in Figure \ref{fig:protocol_deform} and are taken for further investigations.
These include $\lambda = 1.0$ (undeformed), $1.5, ~2.0, ~\text{and}~3.0$.

After the crosslinking and deformation processes, the fourth step in our protocol involves relaxing the system in the NVT ensemble at the same temperature ($T= 0.90$), using a maximum of $100\times 10^6$ MD steps.
In this step, we wait until all pressure components $P_{xx}$, $P_{yy}$, and $P_{zz}$ have reached the expected plateau.
We refer to the diagonal of the pressure tensor calculation provided in the LAMMPS package, which is based on the sum of the kinetic energy tensor and virial tensor.
The read-out of the final pressure values (see Figure \ref{fig:nvt}) will be used as a basis for the next step.

In the fifth step, we perform a relaxation in the NPT ensemble.
Here, we denote the ensemble as \enquote{controlled NPT} where the pressure in the transversal directions, i.e. converged $P_{yy}$ and $P_{zz}$ from the previous NVT step, is maintained by the Nose-Hoover barostat.
This is achieved while simultaneously fixing the extension in the x-direction with $L_x=\lambda L_{x,0}$.
By these boundary conditions, we effectively mimic the experimental conditions of a polymer material under fixed uniaxial elongation with controlled pressure in the transversal directions.
As we have set the conditions from the read-out of the previous NVT step, we observe that even in the NPT step, the pressure $P_{xx}$ stays stable (see Figure \ref{fig:npt}) on the typical time scales of the later cooling and heating runs.
The controlled NPT relaxation is crucial for implementing experimentally relevant conditions in the simulation, where substantial volume changes due to crystallization and melting take place.
We perform the controlled NPT relaxation for $50\times 10^6$ MD steps in the melt state.

The final steps of our protocol involve the cooling (sixth step) and reheating (seventh step) of the system, both of which are carried out in the controlled NPT ensemble (as in the fifth step).
The cooling process begins at a temperature of $T= 0.9$ and is gradually reduced to $T= 0.75 ~(~\widehat{=}~412.5\mathrm{K})$ over the course of $75\times10^6$ steps.
This corresponds to a cooling rate of $2\times 10^{-7}\tau^{-1}$.
The system is then reheated from $T= 0.75$ to $T= 0.95 ~(~\widehat{=}~522.5\mathrm{K})$ in $100\times10^6$ steps, using the same rate as the cooling process.
Please refer to the SI, Section~\ref{sec:protocol} for the illustrations of the relaxation runs, cooling, and heating cycles.

\subsection{True stress}\label{sec:shear_modulus}

\begin{figure}[!ht]
  \subfigure[]
    {
    \includegraphics[width=0.45\hsize]{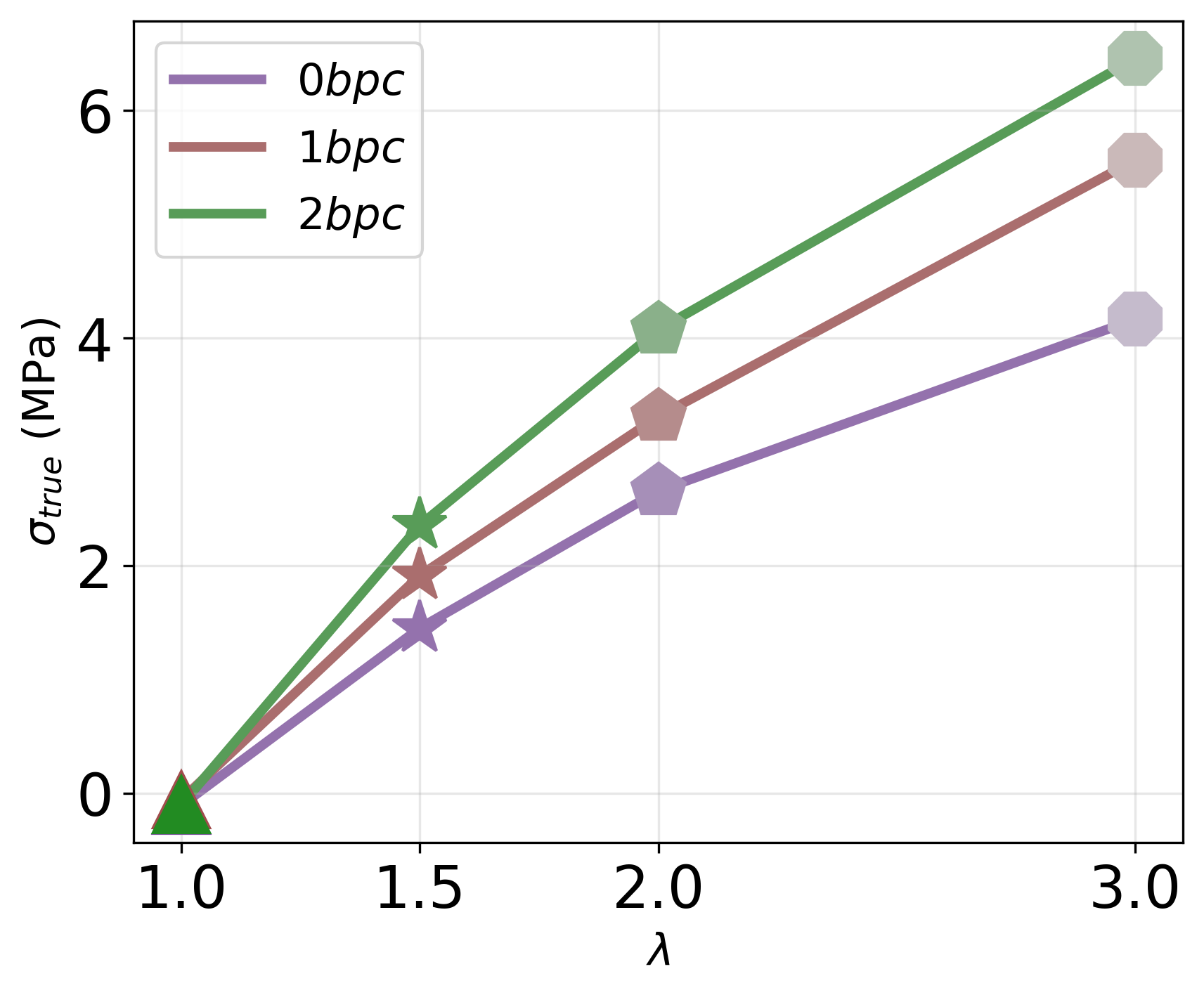}
    \label{fig:stress}
    }
  \subfigure[]
    {
    \includegraphics[width=0.45\hsize]{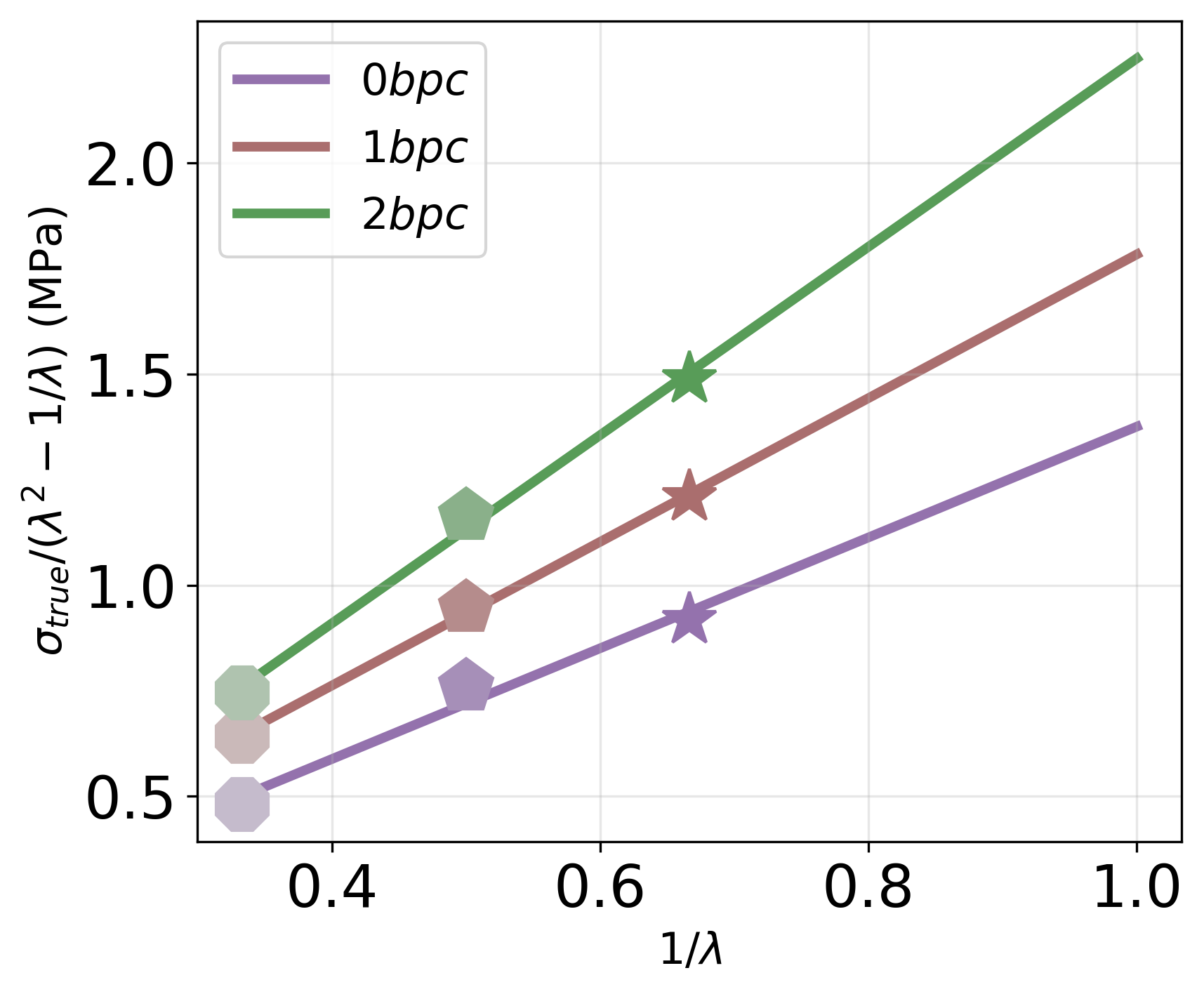}
    \label{fig:modulus}
    }
  \caption{(a) True stress ($\sigma_{true}$) and (b) the Mooney-Rivlin plot at high temperature $T=0.9$ as a function of $\lambda$ and $1/\lambda$, respectively.
  We see the contribution of pure entanglements and deformation in the uncrosslinked system ($0bpc$), and the additional contribution of the network part in the cases of $1bpc$ and $2bpc$.
  The values of Mooney constants are: $C_1 = [0.031,~ 0.040, ~0.008]$ and $C_2= [0.656,~0.851,~1.115]$ for $0bpc$, $1bpc$, and $2bpc$, respectively.
  }
  \label{fig:stress_mod}
\end{figure}

From the ratio of the deformation force to the cross-sectional area of its application, one defines the tensile stress perpendicular to the cross-sectional area and along the direction of the applied force.
For instance, stress along the x-axis ($\sigma_{xx}$) is force ($f_x$) per unit area vector along the x-axis given by $L_y \times L_z$~\cite{rubinstein2003polymer}.
For the other two directions, $\sigma_{yy}$ and $\sigma_{zz}$ are defined equivalently.

The tensile stresses can be measured in terms of the pressure components of the stress tensor wherein negative pressure along the x-axis, $-P_{xx}$ corresponds to $\sigma_{xx}$, $-P_{yy}$ to $\sigma_{yy}$ and $-P_{zz}$ to $\sigma_{zz}$.
Then the true stress ($\sigma_{true}$), is defined as:
\begin{equation}
\sigma_{true} = \sigma_{xx} - \frac{\sigma_{yy}+\sigma_{zz}}{2}~~.
\label{eq:stress_diff}
\end{equation}
Figure~\ref{fig:stress}, shows the true stress at high temperatures (elastic state) due to the applied deformation $\lambda>1$ in the system.
For the undeformed system ($\lambda=1.0$), the value of $\sigma_{true}$ is observed near zero as expected, regardless of network density.
We note that the behavior of the uncrosslinked system is determined by entanglements only as our simulation time is below the terminal relaxation time of the highly entangled melt.
To describe the non-linear stress-strain behavior in the presence of entanglements going beyond the neo-Hookean law, we fit the Mooney-Rivlin model to our simulation results~\cite{rubinstein2003polymer, treloar1975physics, flory1985molecular},
\begin{equation}
\frac{\sigma_{true}}{\lambda^2-\frac{1}{\lambda}} = 2 ( C_1 + \frac{C_2}{\lambda})~~.
\label{eq:modulus}
\end{equation}
In Figure \ref{fig:modulus}, we show the Mooney-Rivlin plots for different network densities against the inverse of deformation along the x-axis.
These lines are fitted with the constants $C_1$ (intercept) and $C_2$ (slope) from Equation~\ref{eq:modulus} and extended towards the case $\lambda = 1$.
For $0bpc$, we see a pure contribution of the entangled strands without a crosslink network formation.
However, as the network density rises in the case of $1bpc$ and $2bpc$, we see that total stress grows.

\subsection{Order parameters}

{\bf{Stem Length (SL).}}
Consecutively aligned monomers (or in trans-trans conformation) are referred to as stems.
For each monomer, the bending angle $\theta_i$ is defined as the angle between the bond vectors of consecutive monomer pairs: ($i-1, i$) and ($i, i+1$), where $i$ is a monomer's index along the chain. A monomer is considered part of a stem if its bending angle satisfies $\theta_i \geq 150 \degree$.

{\bf{Crystallinity ($X$).}}
The fraction of crystalline monomers, $X$, is calculated by counting the number of consecutively aligned monomers in consecutive $trans$-states with a stem length $d_{tt}\geq18$. The threshold $d_{tt}=18$ had been defined earlier with the help of semi-supervised learning-based classification of monomer fingerprints describing crystalline order~\cite{bhardwaj2024}, and where the decision boundary was been optimized regarding its specificity for early nuclei.

{\bf{Orientation Order Parameters ($S$).}}
We are interested in the degree of crystalline order in aligmnent with the direction of pre-stretching, and define
\begin{equation}
    S = (3\langle\cos^2 \theta\rangle-1)/2~~,\label{eq:OOP}
\end{equation}
where $\theta$ is the angle between the bond vector of each monomer and the deformation direction, $x$.

\section{Results and Discussion}\label{sec:results}

\begin{figure}
  \centering
    \subfigure[]{
    \includegraphics[width=.31\hsize]{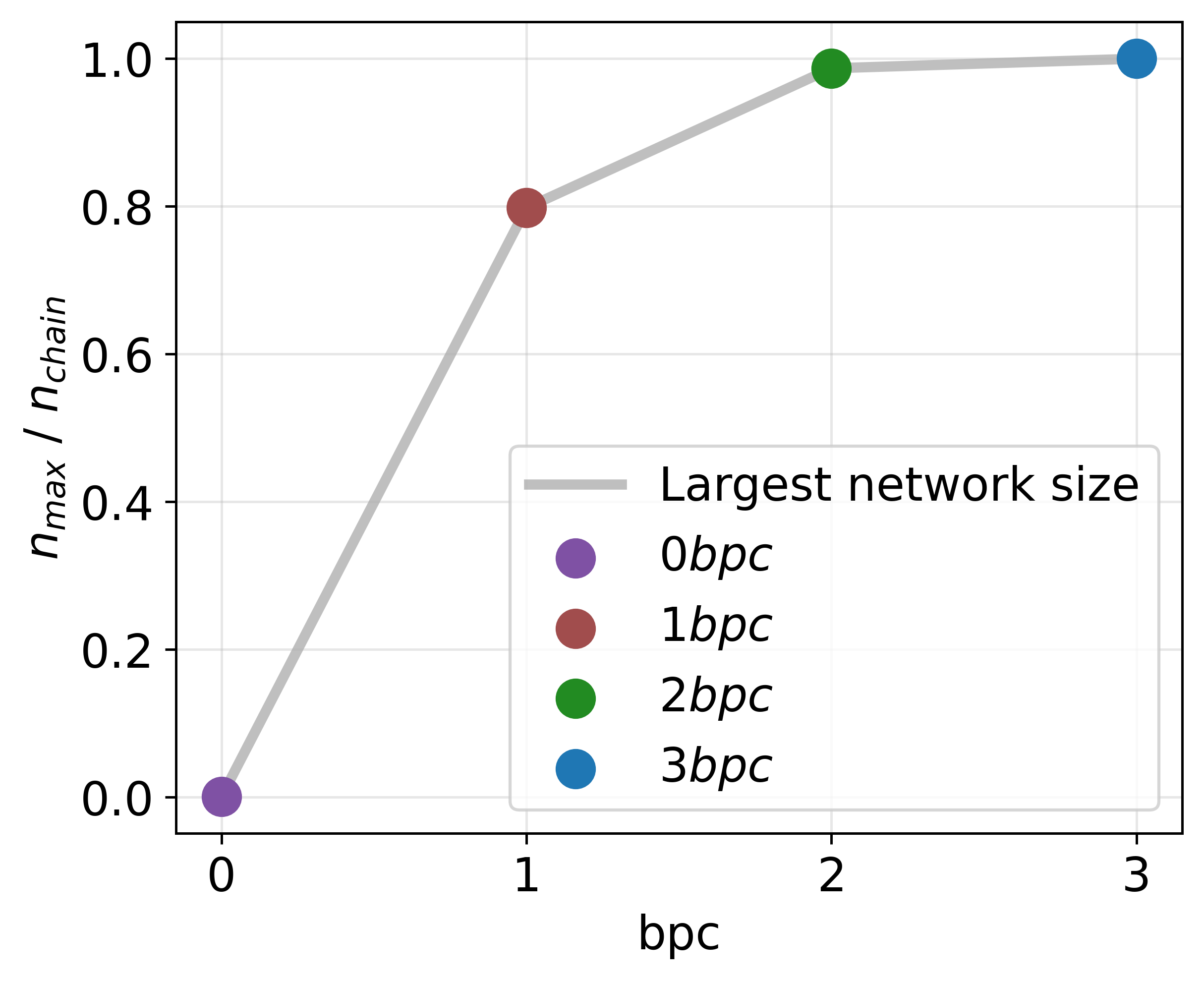}
    \label{fig:max_cluster}
    }
    \subfigure[]
    {
    \includegraphics[width=0.31\hsize]{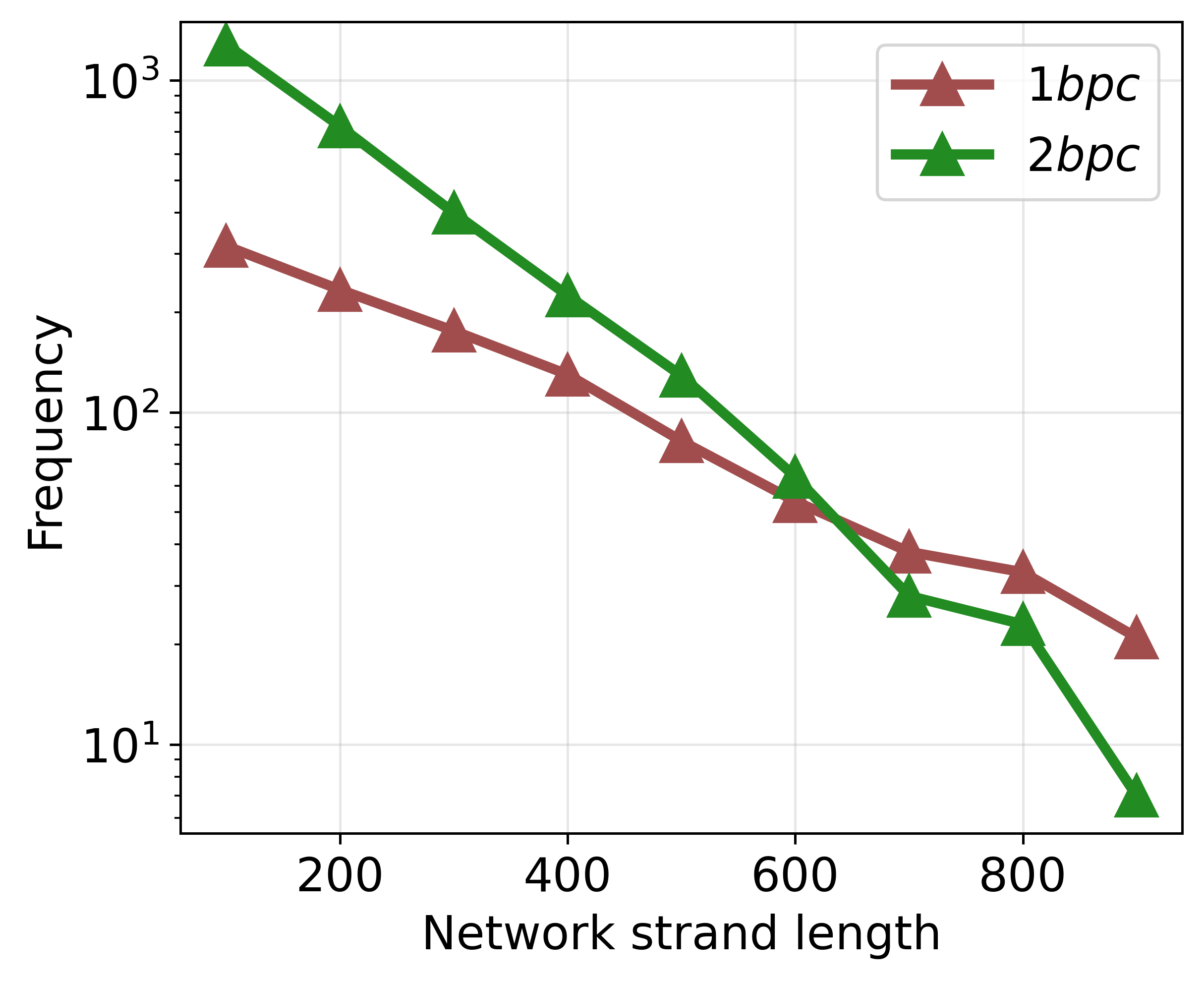}
    \label{fig:active_ends}
    }
    \subfigure[]
    {
    \includegraphics[width=0.31\hsize]{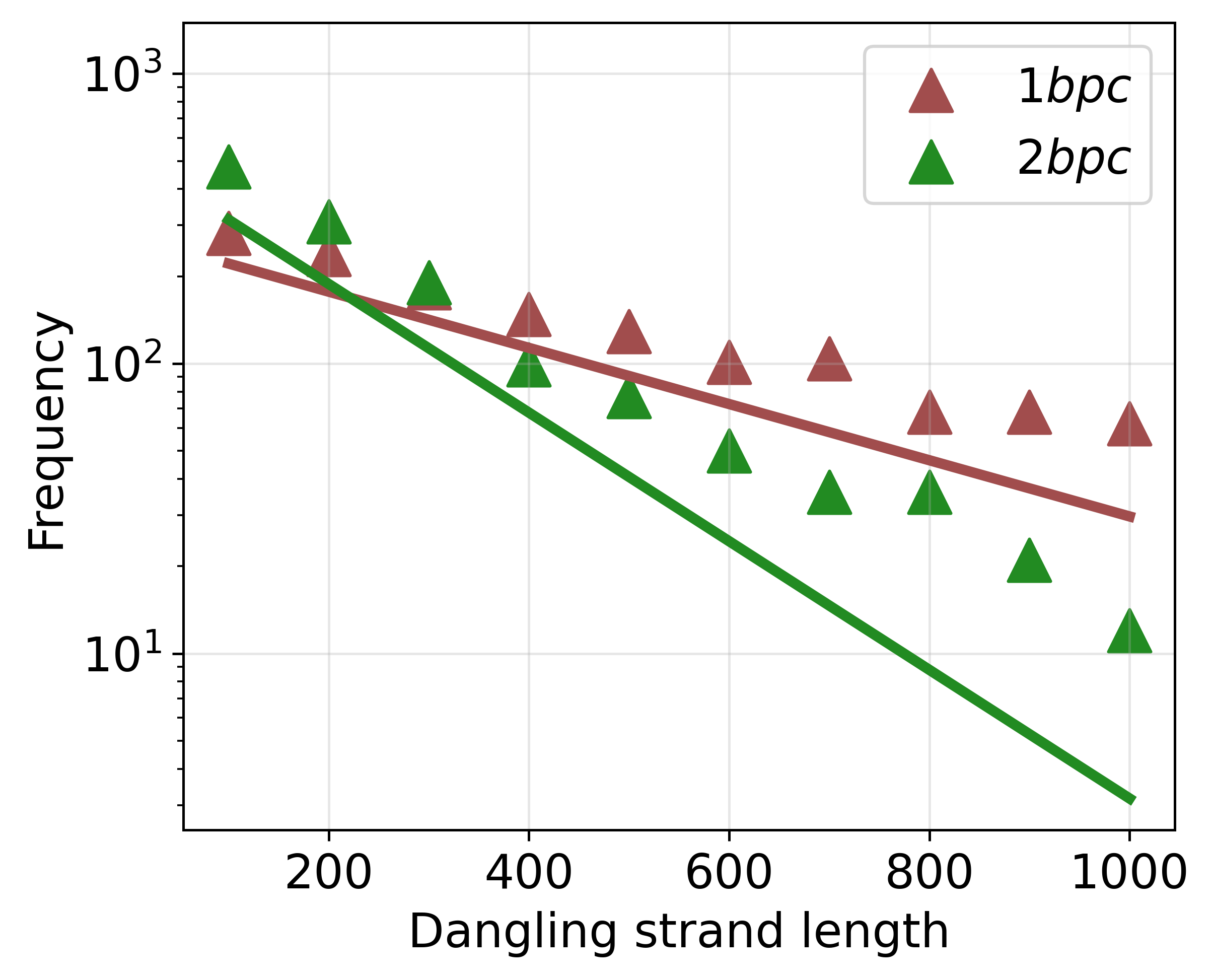}
    \label{fig:dangling_ends}
    }
  \caption{(a) The gel fraction shown as the development in the largest network size ($n_{max}$) of the system with an increasing average number of crosslinks per chain.
  (b) Histogram of elastic network strands, i.e. part of the polymer chain between two crosslinking ends for $1bpc$ and $2bpc$.
  (c) Lengths of dangling strands, i.e. part of the chains with at least one free end.
  The straight lines correspond to the binomial distribution (Equation~\ref{eq:binomial}) of the dangling strands.
  Since the network fills most of the sample as we reach the density of $2bpc$, we do not consider the case of $3bpc$ for further figures in the panel.
  }
  \label{fig:strands}
\end{figure}

In order to evaluate if the network spans the simulation volume, we examine the largest cluster of connected chains~\cite{witte1996} ($n_{max}$).
Figure \ref{fig:max_cluster} illustrates the growth in the gel fraction ($n_{max}/n_{chain}$), as we grow the number of crosslinks in the system.
Notably, we observe that for the case of $1bpc$, approximately $80\%$ of the chains are connected to the largest chain network.
For $2bpc$, more than $95\%$ of the system is a part of the network.
As we reach $3bpc$, all of the chains are connected as a gel.
We note that the theoretical gelation threshold is $0.5bpc$~\cite{rubinstein2003polymer, flory1944network}.

We can identify network strands (or network chains), which refer to the segments of the chain bounded by two crosslinks, and dangling strands, which refer to the uncrosslinked chain ends and uncrosslinked chains.
The length of a network strand is determined by the number of monomers that lie between any two crosslinked monomers.
These strands can contribute to the mechanically active part of the network that sustains external load during deformation in equilibrium.

Given that we have introduced crosslinks in the system, with every $100$th monomer on each chain serving as a crosslinkable (reactive) site, we can expect network strands to range in length from $100$ to $900$.
Only reactive monomers can form an additional bond with each other.
If the first bead of a chain serves as a reactive monomer, two bonds can be formed.
To visualize the distribution of network strand lengths, we plot the results for the cases of $1bpc$ and $2bpc$ in Figure \ref{fig:active_ends}.

In addition, in Figure \ref{fig:dangling_ends}, we illustrate the length distribution of dangling strands in the system.
The distribution of strand lengths is generally given by a
binomial distribution ~\cite{sommer1991, lang2003length} with the probability of finding a  strand of length ($l$) as:
\begin{equation}
    P(l) = p_0~(1-p_0)^l = p_0 \exp({-l/l_0})~~\text{with}~~l_0 = -1/\ln{(1-p_0)}~~.
    \label{eq:binomial}
\end{equation}
Here, $p_0$ is the probability of forming a crosslink, and $l_0$ defines the characteristic strand length, which approaches the average strand length for a small value of $p_0$.
This explains the exponential distribution in Figures~\ref{fig:dangling_ends} and~\ref{fig:active_ends}.
The parameter-free fit for the two values of the crosslink density is indicated in Figure~\ref{fig:active_ends}.

Table~\ref{tab:strand_len} shows the average length of both network and dangling strands for $1bpc$ and $2bpc$.

\begin{table}
  \begin{tabular}{cccc}
    \textbf{Strand type} & \textbf{Network density} & \textbf{Average length} \\
    \midrule
    \multirow{2}{*}{Network}  & $1bpc$ & $295$\\
                              & $2bpc$ & $209$ \\
    \midrule
    \multirow{2}{*}{Dangling} & $1bpc$ & $354$ \\
                              & $2bpc$ & $185$\\
    \bottomrule
  \end{tabular}
  \caption{Summary of average strand lengths from the simulation.
  }
  \label{tab:strand_len}
\end{table}

\subsection{Definition of the characteristic temperatures}\label{si:char_temp}

\begin{figure}[!ht]
  \centering
   \subfigure[]{
   \includegraphics[width=0.47\hsize]{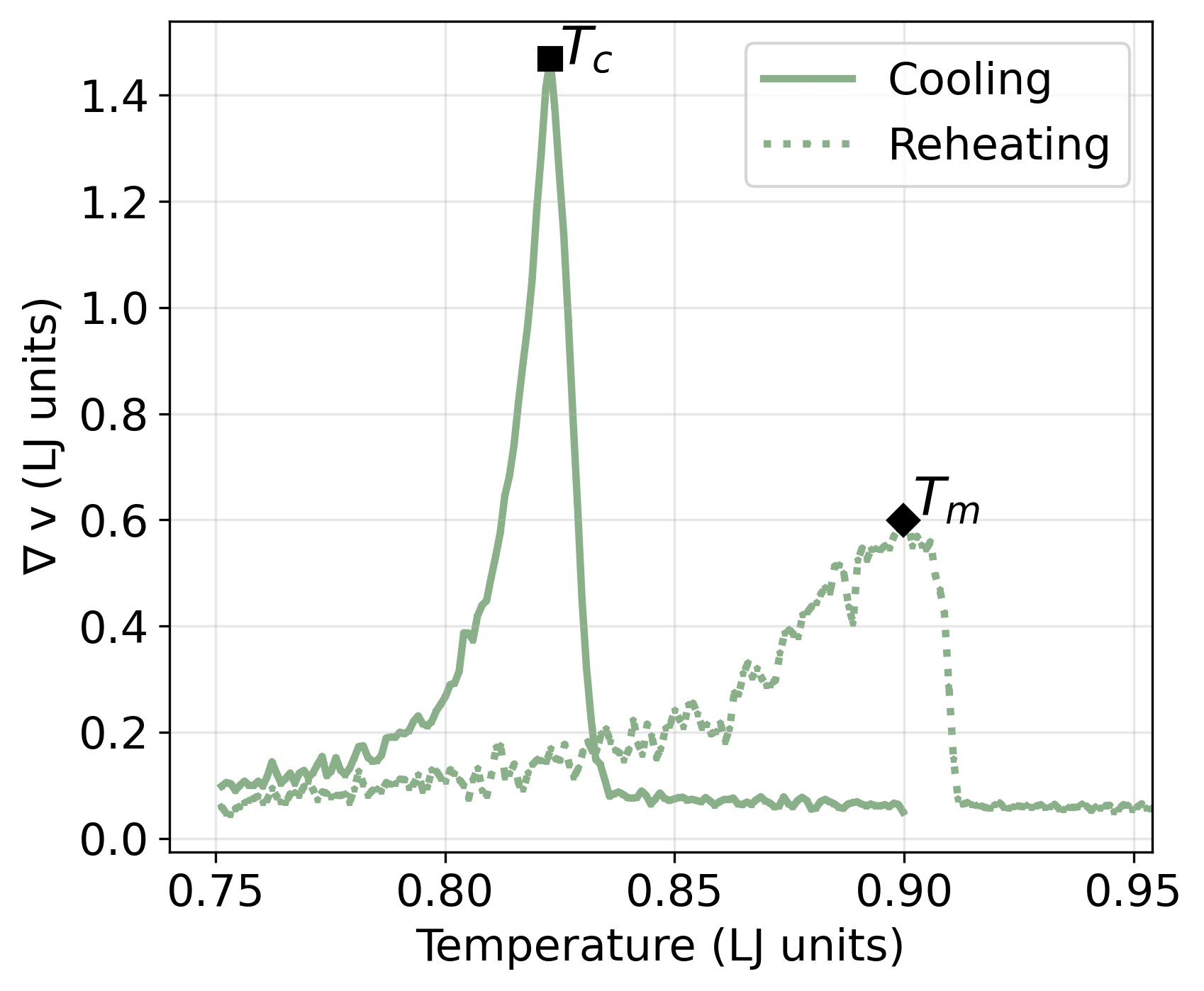}
   \label{fig:del_v}
    }
    \subfigure[]
    {
    \includegraphics[width=0.47\hsize]{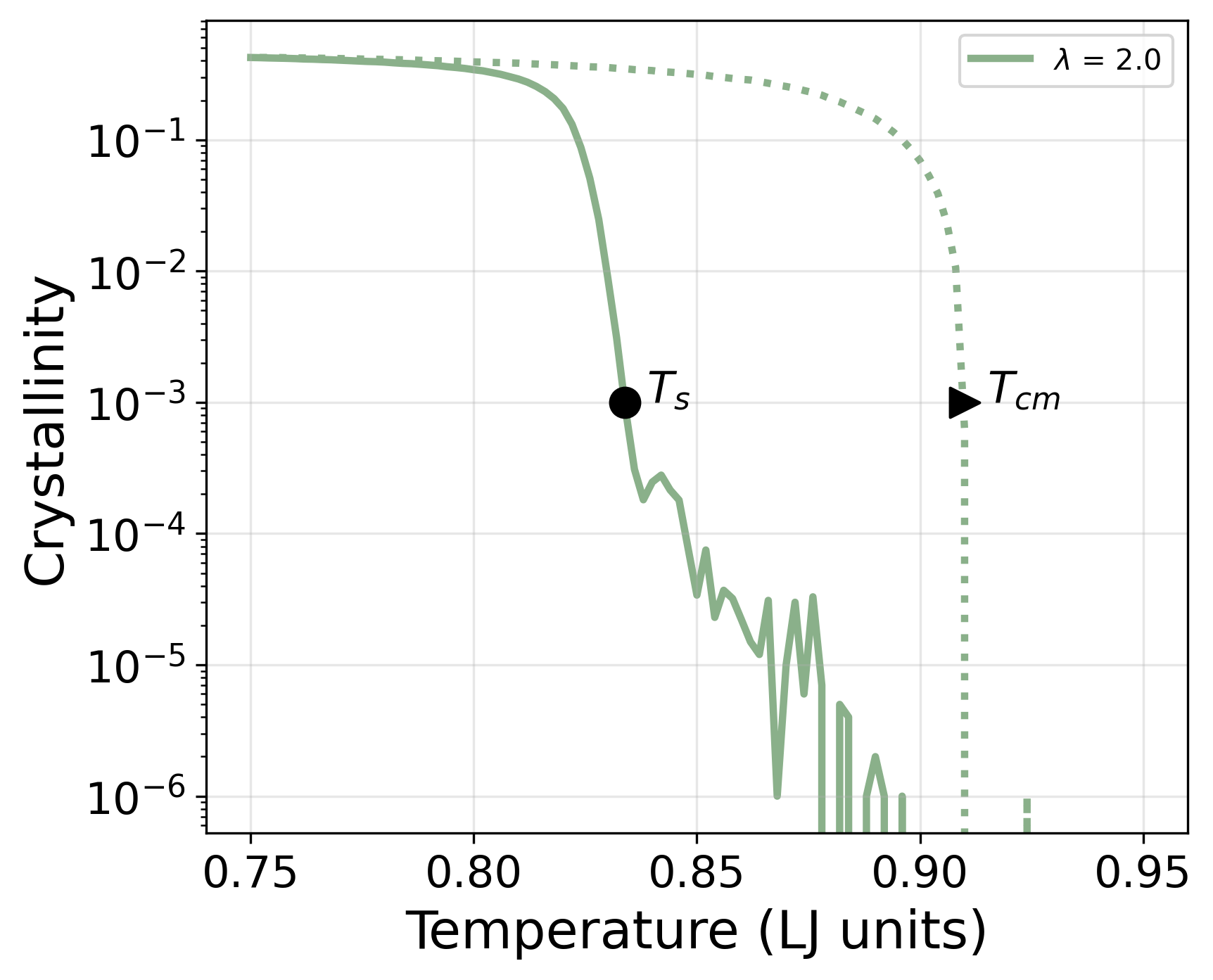}
    \label{fig:crystallinity_cutoff}
    }
   \caption{Definition of characteristic temperatures as used in this work. The graph presents the first derivative of the total volume with respect to temperature (thermal expansion coefficient) during cooling (continuous line) and heating (dotted line) cycles for the undeformed sample $0bpc$.
  }
  \label{fig:def_ch_temps}
\end{figure}

The definition of four characteristic temperatures which are discussed in the following is illustrated in Figure \ref{fig:del_v}. The figure shows the first derivative of the specific volume during cooling and reheating cycles. We defined the crystallization and melting temperatures, $T_{c}$ and $T_{m}$, where the value of $\nabla v$ shows local maxima during the cooling and heating cycles, respectively.
The solidification temperature, $T_{s}$, marks the onset of crystallization, and $T_{cm}$ denotes the temperature of complete melting. 
We define both $T_{s}$ and $T_{cm}$ as the temperatures, where crystallinity $X$ crosses the value of $X=10^{-3}$.
$T_{cm}$ is the complete melting temperature where $\nabla v$ approaches the background slope of $v(T)$, i.e. the thermal expansion coefficient of the molten state. For an overview, see Table  \ref{tab:temperatures}. These four temperatures obey the natural order relation ($T_{c}<T_s <T_{m} <T_{cm}$).
For the cooling cycle, the average of local minima just before the global maximum is taken as the solidification temperature ($T_s$), and the global maximum as the crystallization temperature ($T_c$).
For the heating cycle, the global maximum marks the onset of melting ($T_m$), and the first minimum after $T_m$ represents the complete melting temperature ($T_{cm})$.

\begin{table}
  \begin{tabular}{cccc}
    \textbf{Cycle} & \textbf{Description} & \textbf{Acronym} & \textbf{Symbol} \\
    \midrule
    \multirow{2}{*}{Cooling} & Solidification & $T_s$ & $\newmoon$\\
    & Crystallization & $T_c$ & $\blacksquare$\\
    \midrule
    \multirow{2}{*}{Reheating} & Melting & $T_m$ &  $\blacklozenge$ \\
    & Complete melting & $T_{cm}$ &  $\RHD$\\
    \bottomrule
  \end{tabular}
  \caption{Representation of characteristic temperatures during cooling and reheating cycles.
  }
  \label{tab:temperatures}
\end{table}

\subsection{Cooling and reheating profiles}

\begin{figure}[!ht]
  \begin{center}
  \includegraphics[width=0.9\hsize]{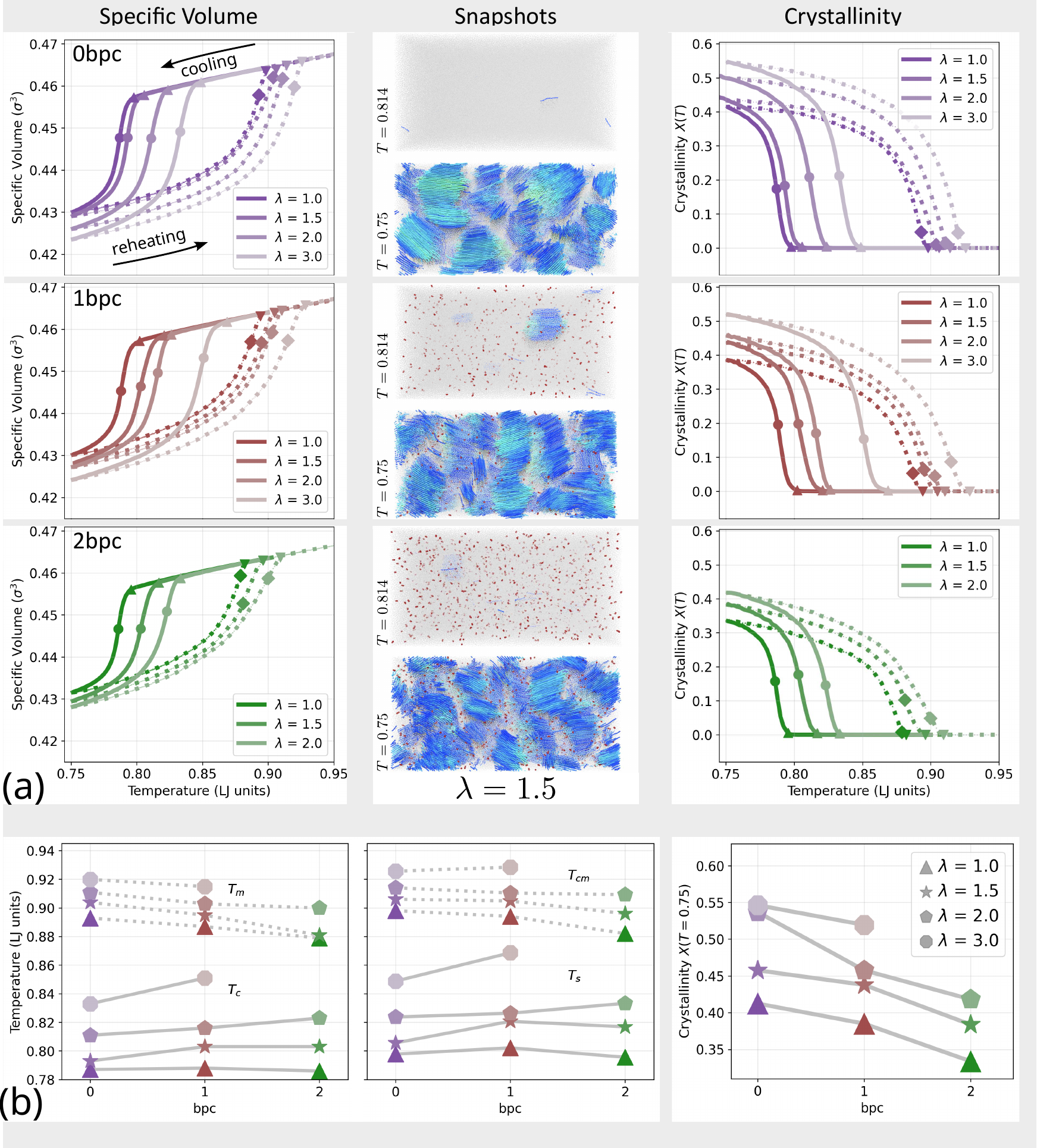}
  \end{center}
\caption{(a) VT phase diagrams for cooling (solid) and heating (dotted) at fixed average crosslinks per chain ($bpc$), shown for pre-stretch ratios $\lambda = 1.0, 1.5, 2.0$, and $3.0$ (Step 3 in Fig.~\ref{fig:flowchart_protocol}). Middle: representative snapshots from the cooling run at $\lambda = 1.5$ and $T = 0.814$ and $0.75$. Amorphous (grey), crosslinked monomers (darg red), and stems (blue shortest, cyan longest) stems are shown. Right: crystallinity ($X$) during cooling and heating for different $\lambda$ at fixed $bpc$.
(b) Dependence of melting ($T_m$), crystallization ($T_c$), complete melting ($T_{cm}$), and solidification onset ($T_s$) on $bpc$ and pre-stretch ratio. Right: crystallinity $X$ at $T = 0.75$ (LJ units) as a function of $bpc$ and $\lambda$.
  }
  \label{fig:vt_x}
\end{figure}

In Figure \ref{fig:vt_x}, we display the specific volume $v$ (Figure \ref{fig:vt_x}(a; left)) and the crystallinity $X$ (Figure \ref{fig:vt_x}(a; right)) as a function of temperature during cooling and reheating for all combinations of the deformation ratio, $\lambda$, and crosslink densities, as well as a summary of characteristic temperatures and maximum crystallinity in Figure \ref{fig:vt_x}(b). The evolution of $v$ and $X$ is shown during the cooling by continuous lines and heating by dotted lines. We use different opacities (see legend) of the same color to distinguish different deformation ratios for a particular crosslinking density.
During the cooling and reheating cycles, we observe the characteristic thermal hysteresis of polymer crystallization, which is far from thermodynamic equilibrium.

To provide a visual representation of the simulation results, we show snapshots of the simulation box at two different temperatures, $T=0.75$ and $T = 0.814$, from the cooling curve for the case $\lambda=1.5$.

For all combinations of $bpc$ and $\lambda$, we observe an abrupt/sharp reduction of the specific volume during cooling (Figure \ref{fig:vt_x}(a; left)). This characteristic behavior is indicative of the discontinuous transition as expected from an under-cooled melt. Here,
$T_{s}$ marks the onset of this phase change, beyond which the crystalline and amorphous phases coexist, giving rise to an apparent semi-crystalline polymer. The density jump during crystallization is in the order of $1/10$ in good agreement with the experimental observations. 

During heating, it can be observed in Ref.~\cite{CH_md2010} that the short stems begin to melt and the longer stems can reorganize to achieve even longer lengths.
As a result, the crystal domains melt and grow simultaneously, resulting in a smoother and more gradual transition and a higher $T_m$. The normal expansion of the molten system is reached again at $T_{cm}$, see also Figure~\ref{fig:del_v}.

\subsubsection*{Effect of deformation}

Upon deformation, the cooling and reheating profiles shift considerably towards higher temperatures.
An upward shift is clearly visible for all characteristic temperatures in Figure \ref{fig:vt_x}(b; left and middle) as we increase the value of $\lambda$.
The earlier onset of crystallization is related to a decrease in the kinetic barrier for nucleation ~\cite{cui2018multiscale}, which may be related to a higher degree of segmental order. Furthermore, the degree of crystalline order in the semi-crystalline state at lowest temperature $T=0.75$ is increased as a result of deformation, which is reflected in the enhancement of $X$ with deformation in Figure \ref{fig:vt_x}(b; right).

Results for the highest crosslink density ($2bpc$) and strong deformation ($\lambda=3.0$) have been omitted. These showed indications of strain-induced crystallization (SIC) which is a known effect upon the stretching of rubber materials~\cite{mark:pes:1979, tosaka2007strain, huneau:rcat:2011}, and will be discussed separately.

\subsubsection*{Effect of crosslinks}

In Figure \ref{fig:vt_x}(b; left and middle part), we display the characteristic temperatures under variation of the crosslink density.
For undeformed systems, both $T_s$ and $T_c$ remain unchanged as the crosslink density rises. This indicates that the weak crosslinking of the highly entangled melt had only marginal influence on the nucleation and growth of the crystalline phase.
In contrast, the values of $T_s$ and $T_c$ in deformed systems increase with crosslink density (Figure \ref{fig:vt_x}(b; left and middle)).
Hence, close to the nucleation time, the formation of crystalline order seems to be promoted by the presence of crosslinks - in the case of stretching. The earlier emergence of crystalline order in the case of crosslinked polymers under stretching is emphasized when following the vertical line in Figure \ref{fig:vt_x}(a; right) that is at $T_c$ for $0bpc$. When introducing crosslinks, the transition shifts to a higher temperature, and crystallinity increases at an equivalent temperature shortly after nucleation.

In contrast, during growth, we see a reduction of crystallinity $X$ with increasing crosslink density and a corresponding decrease of the final value of crystallinity in Figure \ref{fig:vt_x}(b; right).
The reduction of crystallinity in semi-crystalline polymer networks is discussed in the context of additional conformation constraints~\cite{nilsson2010structural} compared to uncrosslinked polymers ($0bpc$).
We emphasize that our results refer to relatively large stretching and cooling rates, challenging a direct comparison of the effects at nucleation and growth stages to the published experimental data.

The retention of the maximally achieved crystallinity $X$ in the presence of crosslinks is followed by faster melting during the subsequent heating cycle.
This is evident, for instance, when we concentrate on the case of $\lambda = 2.0$ across the values of all plots as shown in Figure \ref{fig:vt_x}(a; right).
Also, we see a shift of the decay in $X$ upon heating towards lower temperatures.

\subsection{Cooling and Reheating under Constant Stress}\label{sec:Ensemble_P}

Complementary to the results discussed on the basis of constant strain ($\{L_x,P_{yy},P_{zz}\}$ conditions) above, we performed cooling and heating cycles (Steps 6 and 7 in Fig.~\ref{fig:flowchart_protocol}, respectively) in the case of constant stress, i.e. under given $\{P_{xx},P_{yy},P_{zz}\}$. 

To this end, the configurations resulting from Step 5 (relaxed samples after deformation) are taken as a starting point for each parameter pair (bpc,$\lambda$), and the principal pressure components $\{P_{xx},P_{yy},P_{zz}\}$ were computed from the last $10^6$ MD steps of Step 5. In the subsequent cooling and reheating cycle (Steps 6 and 7 in Figure~\ref{fig:flowchart_protocol}), these values are set as the reference for the Nose-Hoover barostat. We confirmed that without temperature change the box geometry is stable near the deformation state $\lambda$ defined by the pre-stretching.

\begin{figure}[!ht]
  \begin{center}
  \includegraphics[width=\hsize]{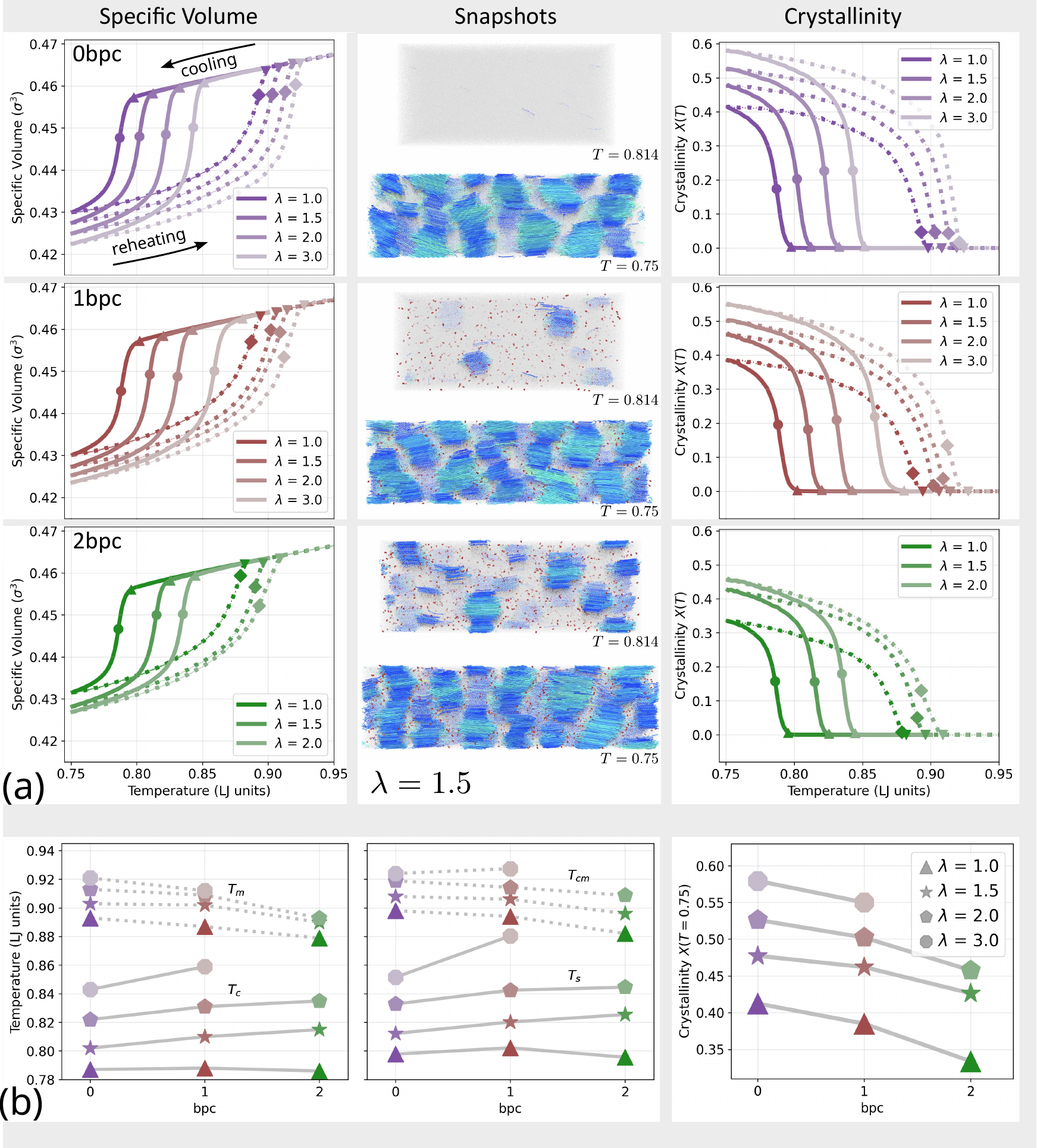}
  \end{center}
  \caption{Analogous presentation of specific volume, crystallinity, and characteristic temperatures as in Figure \ref{fig:vt_x}, but under the constant-stress condition. 
  In (b), we show results under constant stress except in the case $\lambda=1$, for which we repeat the results for constant-strain condition. The legend on the rhs. refers to all three subplots.
  }
  \label{fig:vt_x_ppp}
\end{figure}

In Figure~\ref{fig:vt_x_ppp}, we show the simulation results for specific volume and crystallinity during cooling and reheating under constant-stress conditions $\{P_{xx},P_{yy},P_{zz}\}$ analog to Figure~\ref{fig:vt_x}. By comparison between Figures~\ref{fig:vt_x} and~\ref{fig:vt_x_ppp} we see that specific volume and crystallinity show an equivalent hysteresis pattern as in case of constant strain depending on $bpc$ and pre-stretching deformation ratio $\lambda$. In Figure~\ref{fig:vt_x_ppp}(b), the pre-stretching induced effect of crosslinks leading to higher nucleation temperatures, and inhibited saturation of crystallinity $X$ at lowest $T$ are visible equivalently to constant strain conditions (Figure~\ref{fig:vt_x}(b)). The final degree of crystallinity as shown in Figure~\ref{fig:vt_x_ppp}(b, rhs.) is slightly larger as compared Figure~\ref{fig:vt_x_ppp}(b, rhs.). The thermodynamic driving forces for the transition therefore seem only marginally affected by the different boundary conditions. 

A marked difference between Figure~\ref{fig:vt_x}(a) and Figure~\ref{fig:vt_x_ppp}(a) is, however, visible regarding the simulation snapshots in terms of the final box geometry, and the apparent degree of orientation order in the pre-stretching direction. In order to characterize the shape change, we consider the deformation factor during cooling and re-heating. 
The temperature dependent deformation factor, $\lambda(T)$, is calculated as
\[
 \lambda(T) = \frac{1}{2}\left[\left(\frac{L_x}{L_y}\right)^{2/3}+\left(\frac{L_x}{L_z}\right)^{2/3}\right]~~,\label{eq:lambda_T}
\]
reflecting the average resulting from the expected ratios $L_x(T)/L_{y,z}(T)=\lambda(T)^{3/2}$ with respect to both axis $y$ and $z$ perpendicular to the deformation axis. Above expression profits from the statistics in both directions $y,z$, and is independent from specific volume variations that would factor into the ratio $L_x/L_{x,0}$.
In our notation, we distinguish the target deformation for the pre-stretching (Step 3 in Figure~\ref{fig:flowchart_protocol}) denoted as $\lambda$, and the further development of the deformation factor during cooling as $\lambda(T)$.

\begin{figure}[!ht]
  \begin{center}
  \includegraphics[width=\hsize]{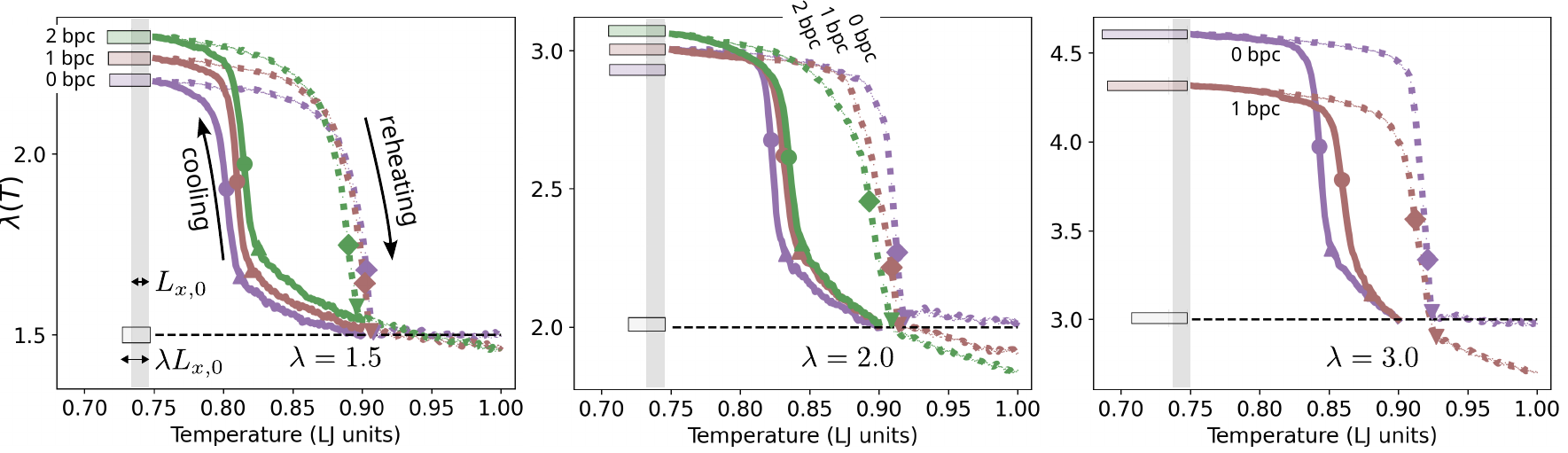}
  \end{center}
  \caption{Evolution of deformation, $\lambda(T)$, under constant-stress conditions during cooling (continuous) and reheating (dashed) after pre-stretch at given $\lambda$ denoted as horizontal dashed lines. The number of cross-links per chain ("bpc") is labeled near correspondingly colored curves. The shapes of the simulation box in $x$ and $y/z$-directions at lowest temperature are illustrated using rectangles of the same color in reference to the pre-stretched condition (grey rectangle) and undeformed box length $L_{x,0}$ (thickness of the vertical grey bar).
  }
  \label{fig:lambda_ppp}
\end{figure}

In Figure~\ref{fig:lambda_ppp}, the simulation results for $\lambda(T)$ are presented for three different degrees of pre-stretching deformation factor $\lambda=1.5,~2.0,~\text{and}~3.0$. For all three values of $\lambda$, the deformation factor increases in the order of $50\%$ upon transition into the semi-crystalline state beside a minor variation of the maximum deformation value $\lambda(T=0.75)$ depending on the number of crosslinks per chain. 

The spontaneous extension of a stretched elastomer along the stretching direction during crystallization appears counterintuitive, as it contrasts with the typical volume contraction observed in most materials upon crystallization. We note that our samples exhibit this volume transition concomitantly with the uniaxial extension, as shown in Fig.~\ref{fig:vt_x_ppp}. At the same time, we reproduce a well-established empirical phenomenon that forms the basis of two-way shape-memory materials \cite{dolynchuk2017reversible,wang:pims:2019,murcia:m:2021}. The physical origin of this behavior is intimately related to the semi-crystalline nature of polymeric materials and is likely associated with an increase in the conformational entropy of the elastic, yet uncrystallized strands during the transition, driven by the spontaneous stretching of crystalline chain segments along the direction of applied stress. We discuss some intriguing aspects of this behavior further below.

\subsection{Comparison of Order Parameters}

\begin{figure}[!ht]
  \begin{center}
  \includegraphics[width=0.8\hsize]{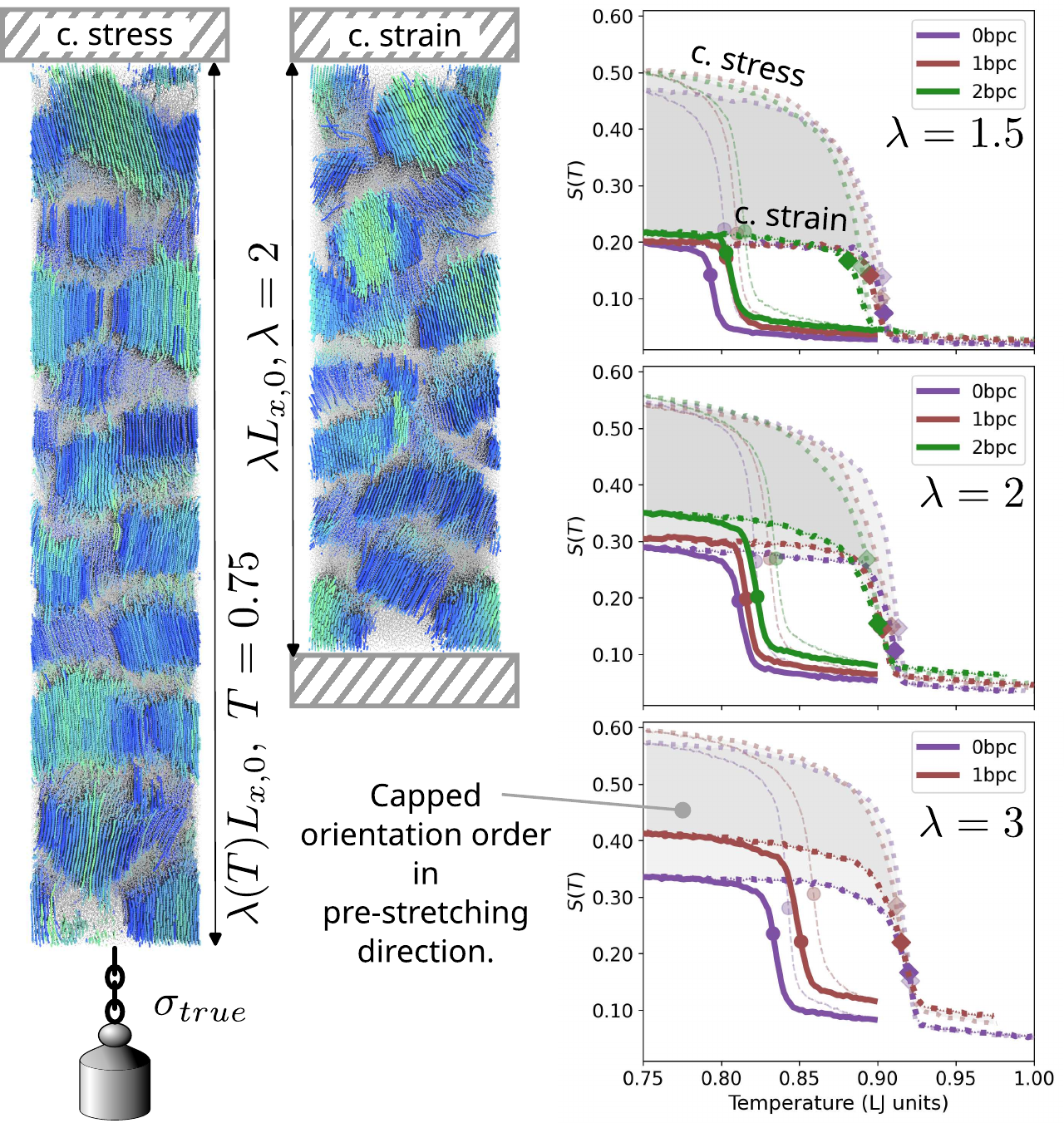}
  \end{center}
  \caption{Orientation order (eq.~(\ref{eq:OOP})) parameter $S(T)$ with respect to the pre-stretching direction $x$ during cooling under constant (c.) stress and strain condition. On the rhs. we show results grouped by given pre-stretching target $\lambda$ (Step 3 in Fig.~\ref{fig:flowchart_protocol}). There, we compare c. strain (dark lines: cooling, dark thick dashed lines: reheating) with c. stress conditions (thin dashed lines: cooling, light thick dashed lines: reheating) for various number of cross-links per chain (bpc), and illustrated the different between both conditions during reheating. Characteristic transition temperatures, $T_c$ and $T_m$ are labeled by circles, and squares, respectively.
  The lhs. compares simulation snapshots for 1bpc and pre-stretching target $\lambda=2$ after cooling under both conditions. Snapshots are colored as in Figure~\ref{fig:vt_x}(a), except omitted crosslinked monomers.
  }
  \label{fig:OOP_lpp_ppp}
\end{figure}

We inspect the role of boundary conditions (constant strain vs. constant stress) on the development of orientation order upon cooling and reheating by means of the orientation order parameter $S(T)$ (eq.~\ref{eq:OOP}) of normalized bond vectors with respect to the pre-stretching direction $x$.
The results are shown in Figure \ref{fig:OOP_lpp_ppp}. In the absence of deformation, $\lambda=1.0$, the system exhibits a random alignment, resulting in $S$ fluctuating around zero. Therefore, results shown in  Figure~\ref{fig:OOP_lpp_ppp} focus on the cases $\lambda>1$.

Upon cooling (reheating) we observe the crystallization-dependent strengthening (decay) of orientation order near the characteristic temperatures $T_c$ and $T_m$ for all combinations of $\lambda$ (pre-stretch) and $bpc$ under both constant strain and constant-stress condition. For both conditions, the degree of alignment increases with increasing pre-stretching deformation, $\lambda$.
However, under constant strain, our results indicate that uniaxial orientation order can not develop to the full degree observed under constant-stress conditions, but is "capped" by a smaller saturation value. 
This is particularly visible for the smallest pre-stretching deformation $\lambda=1.5$: Under constant strain, development of $S(T)$ almost stops near the inclination point of specific volume $T_c$ during cooling, while under constant stress, develops far beyond $T_c$, and reaches maximum values more than $2\times$ larger than in case of constant strain.
Note that under constant strain, crystallinity does still grow below $T_c$ (Figures~\ref{fig:vt_x_ppp}), for instance, for $\lambda=1.5$ from $X\approx 0.2$ to $X> 0.4$ for all values of $bpc$, while $S(T)$ is already close to its final value during cooling. This means that below $T_c$, crystallites continue to grow, but start to rotate out from the preferred director and / or nucleate in different directions (see Figures~\ref{fig:vt_x}(a) center and snapshots in Figure~\ref{fig:OOP_lpp_ppp}).

\begin{figure}
  \centering
    \subfigure[]{
    \includegraphics[width=.31\hsize]{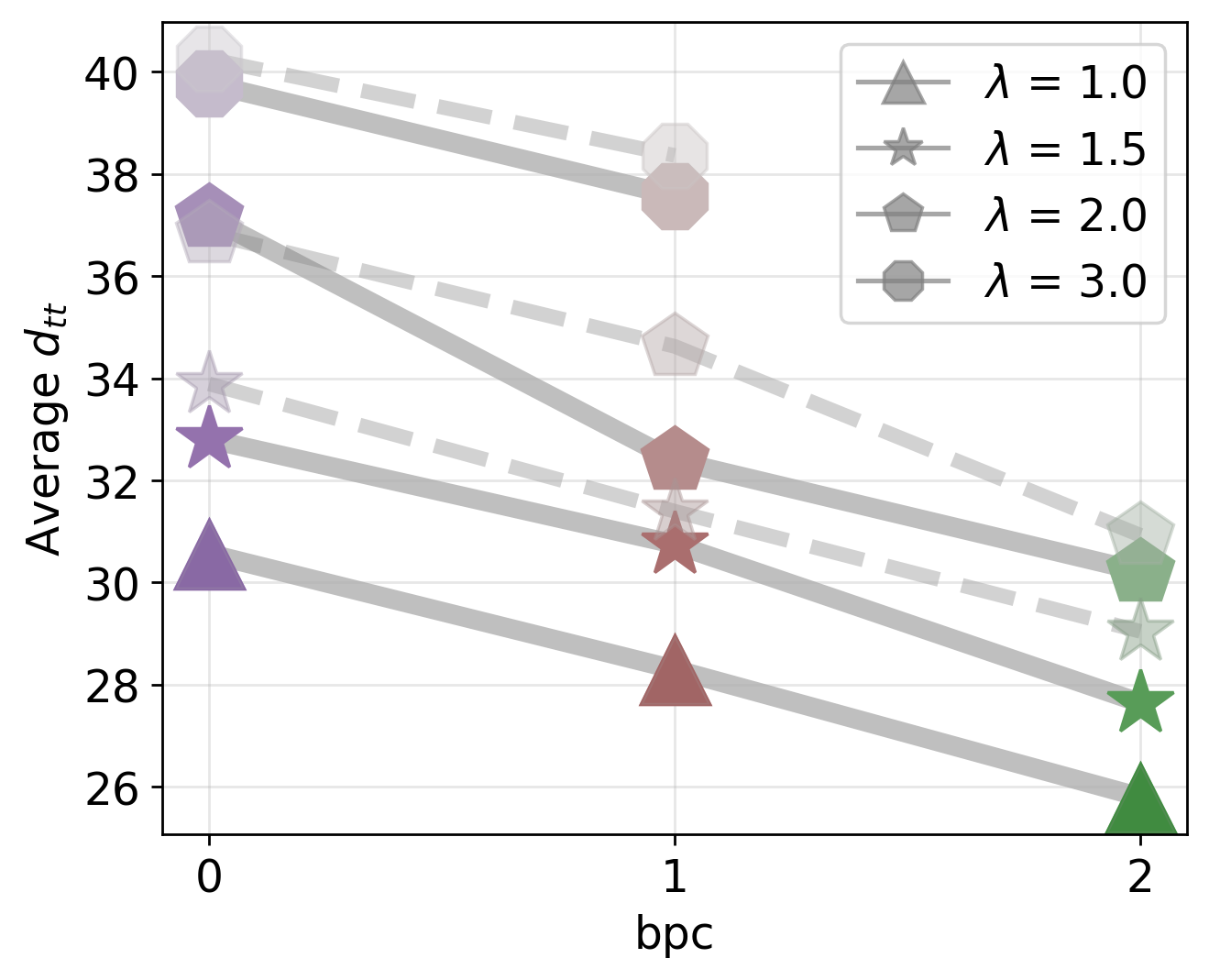}
    \label{fig:avg_sl}
    }
    \subfigure[]{\includegraphics[width=.31\hsize]{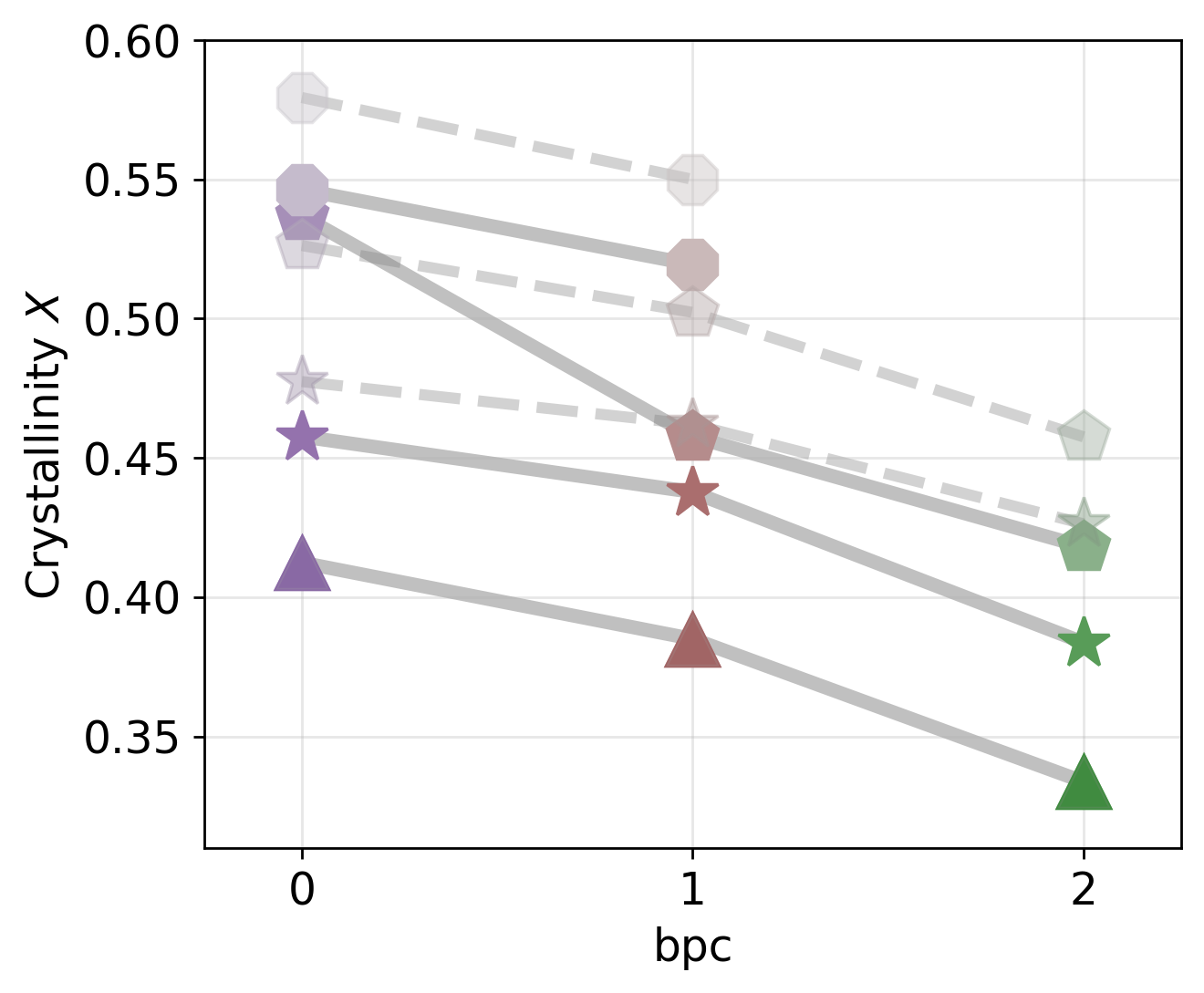}
    \label{fig:cl_X}
    }
    \subfigure[]{\includegraphics[width=.31\hsize]{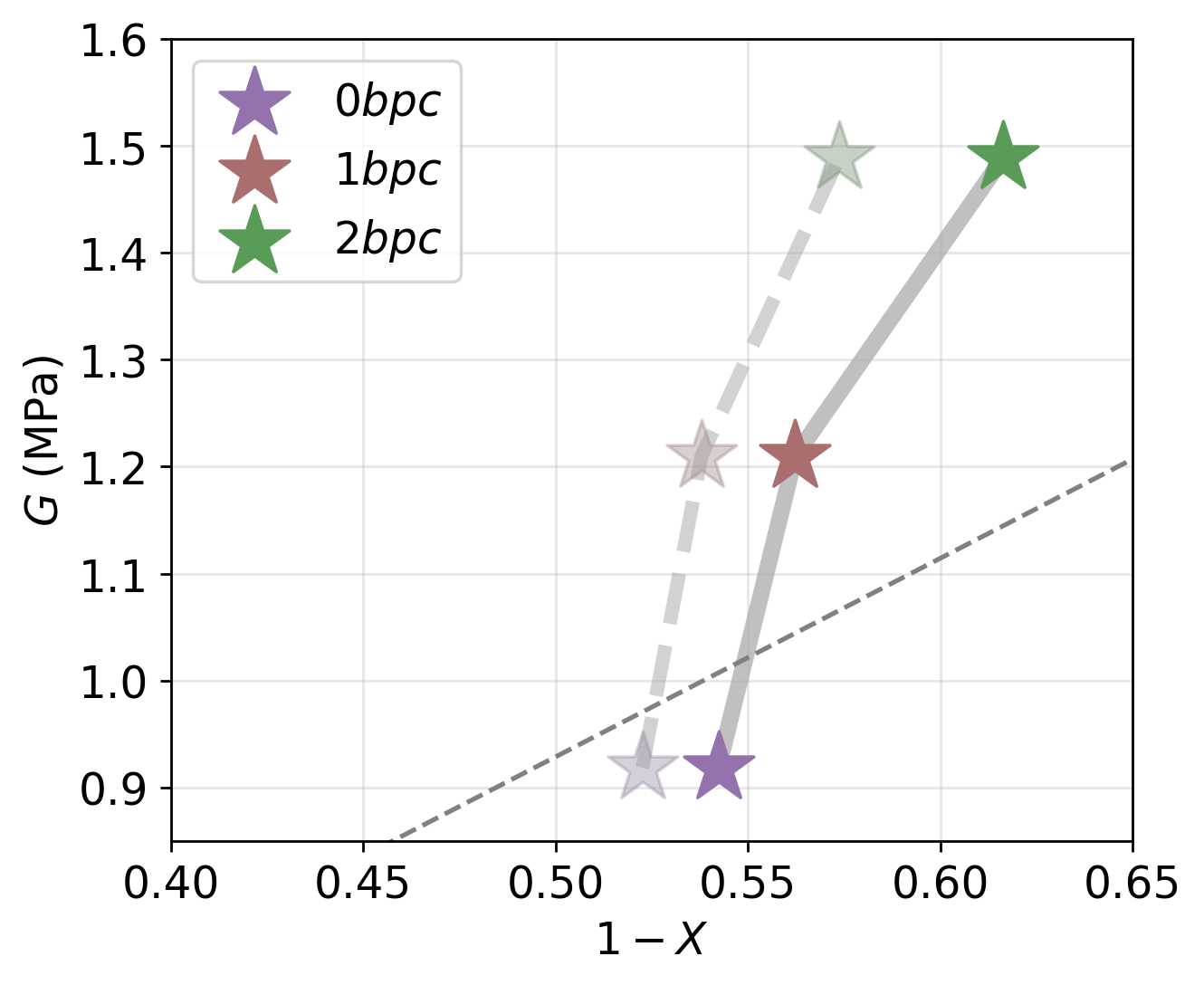}
    \label{fig:amorph_g}
    }
  \caption{At lowest temperature ($T=0.75$): (a) average stem lengths ($d_{tt}$) and (b) crystallinity $X$ under variation of network density $bpc$ and pre-stretching deformation $\lambda$ for constant strain (solid line) and constant stress (dashed line, greyed-out colors).
  (c) Correlation between the amorphous fraction at low temperatures of the undeformed sample with the modulus ($G$) obtained from the affine deformation at $\lambda=1.5$ from Equation~\ref{eq:affine}. The thin dashed line shows a linear curve crossing the origin (zero offset) as a guide to the eye.}
  \label{fig:avg_sl_oop}
\end{figure}

The degree of crystalline order (irrespective of its orientation) shows a small increase when switching from constant strain to constant stress conditions. This is seen in Figure~\ref{fig:avg_sl_oop}, where the average stem length $d_{tt}$ (Figure~\ref{fig:avg_sl}) (number of consecutive trans-trans bond vectors), and the crystallinity $X$ (Figure~\ref{fig:cl_X}) after cooling are compared between the two boundary conditions. There, both values $d_{tt}$ and $X$ are typically larger in case of constant stress as compared to constant strain, where relative differences are in the order of $10\%$ and less.

In summary, the results from and Figures~\ref{fig:OOP_lpp_ppp} and Figure~\ref{fig:cl_X} confirm that pre-stretching defines a thermodynamically preferred crystallite orientation, and demonstrate that the director-compliant crystallite development is less constraint under constant-stress conditions as compared to constant-strain conditions. The additional deformation $\lambda(T)$ in pre-stretching direction under constant stress (Figure~\ref{fig:lambda_ppp}) accompanies a facilitated transition towards the energetically preferred crystalline state. Hence, an excess of internal stress must be present under constant strain below the transition point, which is avoided when crystallization occurs under constant stress.

Both crystalline growth and their freedom of orientation depend on the elastic response of the amorphous phase that is encoded in form of the combination of entanglements and crosslink topology forming the material's memory already at high temperatures. The results shown in Figure~\ref{fig:avg_sl} indicate a monotonous decrease in the average $d_{tt}$ as network density ($bpc$) increases, highlighting the influence of crosslinking on restricting stem-like chain conformations. An increase in network density is expected to lead to a higher number of short stems, while a higher degree of deformation promotes the formation of longer stems for a given network density. Consequently, the longest stems and largest crystallinity (Figure~\ref{fig:cl_X}) are observed in the uncrosslinked system under maximum deformation.

We define the affine shear modulus, $G$, for stretched samples at the high temperature state according to Equation~\ref{eq:modulus} as
\begin{equation}
     G = \frac{\sigma_{true}}{\lambda^2-\frac{1}{\lambda}}~~.
    \label{eq:affine}
\end{equation}
Figure~\ref{fig:amorph_g} shows a correlation between the amorphous fraction of undeformed samples after crystallization and the shear modulus in the amorphous state (obtained from data at $\lambda=1.5$). 

The presentation of Figure~\ref{fig:amorph_g} is inspired by a recent study~\cite{wang2023how} hypothesizing that the minimal thickness of the amorphous layer is determined by a number of entanglements. In case of crosslinked networks this argument might be extended to the sum of all topological constraints including the crosslinks. In simple words: The equilibrium modulus should be directly related with the amorphous fraction.
However, it is an ongoing challenge~\cite{peng2024role}, to theoretically describe on the dependence of the thickness of amorphous layers on the final number of entanglements (i.e. after crystallization) in the system.
In Figure~\ref{fig:amorph_g}, we see that the elastic contribution at high temperature indeed correlates with the amorphous fraction. Absence of a strict proportionality indicates that further investigation is needed to elucidate the relation between the density of topological constraints and the resulting amorphous fraction.

\section{Conclusions}\label{sec:conclusion}

We have conducted large-scale simulations of the crystallization and melting behavior of crosslinked and stretched polymer melts. The starting point was a melt of entangled chains, which also served as a reference system for simulations under stretching. The chosen crosslinking densities were above the gel point but were selected to ensure sufficiently long network strands with an average length of more than 200 coarse-grained monomer units. Both uncrosslinked and crosslinked systems have undergone a rapid uniaxial stretching followed by  a cooling and reheating cycle. The cooling-reheating cycle of all systems has been performed under constant-strain as well as constant-stress boundary conditions in order to compare their effect on crystalline order and mechanical response of the material during the otherwise identical thermal history.

At high temperatures, the considered melts and networks exhibited the expected entropic elasticity of entangled elastomers, and were found in a non-equilibrium state with respect to long-time entanglement relaxation. The deformation ratio was kept below a point of strain-induced crystallization. During the cooling-reheating cycle, the thermodynamic signature by means of specific volume of all simulated systems shows the typical hysteresis known for semi-crystalline polymers with distinct crystallization and melting transition points. Here, the effect of stretching aligns with reported results from the literature in terms of a shift of both crystallization and melting temperatures towards higher values with increasing degree of stretching irrespective of the degree of crosslinking and boundary conditions. 

Without the stretching, crosslinking density has no significant influence on the crystallization transition temperature  (onset of solidification) compared to uncrosslinked melts. From this one can conclude that the nucleation is not noticably affected by crosslinking for the chosen crosslinker densities. In case of stretching, a slight shift towards higher crystallization temperatures in case of crosslinked systems is observed as compared to uncrosslinked systems.
At the lowest temperatures, however, we observe a reduction of stem length and a smaller crystallinity for both undeformed and stretched samples with a higher degree of crosslinking. Upon reheating of these samples, correspondingly, melting points are shifted towards smaller temperatures with increasing number of crosslinks. As a result, a narrowing of the hysteris with respect to temperature is observed by increasing the crosslink density. The results can be related to the interplay between two effects: on one hand, crosslinking reduces the entropy of the pre-stretched network chains, which can accelerate nucleation. On the other hand, the restriction of chain mobility and freedom for reorientation near crosslinking points counteracts folding and growth processes leading to the observed suppression of crystalline order at lowest temperatures.
The resulting impact of crosslinking and stretching on the hysteresis was qualitatively identical under both constant-strain and constant-stress boundary conditions indicating that the thermodynamic driving forces for the phase transitions are equivalent.

In contrast, a marked difference can be observed between the resulting shapes and degree of alignment of crystalline order with stretching direction between constant-strain and constant-stress conditions: under constant stress, a spontaneous elongation of the systems by $~50\%$ beyond the deformation ratio set by the elongation of the molten system is observed for all considered degrees of stretching and crosslink density. This effect is accompanied by a substantial strengthening of crystal orientation towards the stretching direction as compared to the constant-strain boundary conditions. For moderate deformation ratios ($\lambda=1.5$), we see that the orientation order is capped at values near the transition point, while under constant stress, orientation order develops further far below this point. When orientation order saturates under constant strain, crystals continue to develop, but start to become oriented off the preferred director during a later stage of crystallization. This indicates a two-stage crystallization process under the constraint of constant deformation with respect to the development of crystalline morphology.

The shape change upon under cooling under constant stress in our simulations is equivalent to the sudden increase in elongation in experiments on semi-crystalline elastomers under constant stress~\cite{dolynchuk2017reversible,smith:jrnbsn:1938}, and complementary to the experimentally observed stress release during the transition~\cite{murcia:m:2021,gent:tfs:1954}. The resulting temperature-strain hysteresis serves as a starting point for the development of polymer-based shape-memory materials~\cite{murcia:m:2021}. 

We observe that the increase in the amorphous fraction with increasing crosslink density correlates in our simulations with the elastic modulus, which roughly corresponds to the sum of the entanglement and crosslinking densities. This is consistent with the hypothesis of a maximum entanglement (or defect density) in the amorphous fraction, as recently proposed~\cite{wang2023how}. With a preferred director, the elastic response of the amorphous fraction can be linked to a relative entropy increase of the remaining elastically active chains upon crystallization, as stem alignment along the stretch shortens the end-to-end distance of flexible segments, as proposed by Flory~\cite{flory1947thermodynamics}. Large-scale simulations of crystallization of crosslinked melts offer now deeper insights into the physics of shape-memory materials.

In summary, across all degrees of stretching and crosslink densities, the simulated systems show the characteristic hysteresis in the specific volume during cooling and reheating. Stretching  before crystallization  shifts both melting and crystallization to higher temperatures. The influence of crosslinking on the crystallization temperature depends on the degree of stretching. Under constant stress, crystallization induces a spontaneous elongation of the systems on the order of $50\%$ beyond the pre-stretch and a substantial strengthening of orientation order with respect to the stretching direction, whereas under constant strain orientation order saturates while crystalline growth continues. This points to a two-stage crystallization process under constant strain, where in the second stage, the crystalline fraction reorients and / or additional crystalline order develops off-axis.
Overall, the simulated systems reproduce key signatures of shape-memory behavior in semi-crystalline materials. We hope that our simulation results inspire efforts for a better quantitative understanding and the physics-informed tailoring of shape-memory materials.


\begin{acknowledgement}

J.U.S. acknowledges financial support from the German Research Foundation (Grant No. SO 277/19).
The authors acknowledge the Center for High-Performance Computing (ZIH) Dresden for granting simulation time. We also thank Ankush Checkervarty for the technical support in performing the simulations and Dr. Michael Lang for his help in classifying the network structure.
OpenAI ChatGPT (models 5.1 and 5.2) was used to assist in (i) optimizing LAMMPS simulation input/workflow scripts, (ii) drafting Python scripts for processing simulation log files and compressing simulation trajectories into HDF5 files, and (iii) refactoring and optimizing hand-written plotting scripts for order parameters into a unified plotting interface. All AI-assisted code and outputs were reviewed and validated by the authors, and compared against pre-existing hand-written implementations.

\end{acknowledgement}

\begin{suppinfo}

We provide simulation details and further comparisons within the network densities with constant deformation ratio, which are available free of charge as suppinfo.pdf: Supporting Information.

\end{suppinfo}

\bibliography{references,NR} 

\end{document}